\documentclass[a4paper,11pt]{article}
\pdfoutput=1

\usepackage{jheppub}
\usepackage{xcolor,colortbl}
\allowdisplaybreaks

\usepackage{physics}
\usepackage{slashed}
\usepackage{subfigure}

\def\MSbar{\overline{\mathrm{MS}}}

\def\ep{\epsilon}
\def\g{\gamma}
\def\z#1{{\zeta_{#1}}}
\def\ca{{C^{}_A}}
\def\cf{{C^{}_F}}
\def\tf{{T^{}_F}}
\def\nf{{n^{}_f}}
\def\nl{{n^{}_l}}
\def\nh{{n^{}_h}}

\def\as{{\alpha_s}}
\def\asb{{\alpha_s^0}}
\def\asnf{{\alpha_s^{(\nf)}}}
\def\asnl{{\alpha_s^{(\nl)}}}

\def\lmu{l_\mu}
\def\lmu#1{l^{#1}_\mu}

\def\lb#1{\if 1#1 \ln\beta \else \ln^#1\beta \fi}
\def\lt#1{\if 1#1 \ln 2 \else \ln^#1 2 \fi}

\title{Polarized double-virtual amplitudes for heavy-quark
pair production}

\author[a]{L. Chen,}
\author[b]{M. Czakon}
\author[b]{and R. Poncelet}
\affiliation[a]{Werner-Heisenberg-Institut, Theoretical Physics
  Division, Max-Planck Institute for Physics, \\ D-80805 M\"unchen,
  Germany}
\affiliation[b]{Institute for Theoretical Particle Physics and Cosmology, 
RWTH Aachen University, \\ D-52056 Aachen, Germany}
\emailAdd{longchen@mpp.mpg.de}
\emailAdd{mczakon@physik.rwth-aachen.de}
\emailAdd{poncelet@physik.rwth-aachen.de}
\abstract{
 We present the two-loop virtual amplitudes for heavy-quark pair production
 in light quark-antiquark annihilation and gluon fusion channels, 
 including full spin and color dependence. We use expansions around kinematical
 limits and numerical integration to obtain results for the involved master
 integrals. From these, we determine the renormalised infrared 
 finite remainders of the
 coefficients of amplitude decompositions in terms of color and
 spin structures.
 The remainders are given in form of numerical interpolation grids supported
 by expansions
 around the production threshold and the high energy limit.
 Finally, we provide the spin density matrix, which encodes the
 heavy-quark spin correlations and is sufficient for phenomenological
 applications. Our results are necessary for the derivation of
 top-quark pair production cross sections in hadron collisions in the
 narrow width approximation with next-to-next-to-leading order
 accuracy in QCD.
}

\dedicated{\rm MPP-2017-268, TTK-17-49}

\begin{document} 
\maketitle
\flushbottom

\section{Introduction}
The top-quark is the heaviest known particle and measurements of its properties
provide important insights into the Standard Model of Particle Physics
and beyond. 
Top-quark pair production at hadron colliders like the LHC or Tevatron is an important
process for Standard Model precision measurements as well as searches for new
physics.
Considering the hadronic production of stable top-quark pairs,
the prediction from Quantum Chromodynamics (QCD) is compelete to
next-to-next-to leading order (NNLO) for the total cross section
\cite{Czakon:2013goa} and for differential distributions
\cite{Czakon:2015owf,Czakon:2016ckf, Czakon:2016dgf}.
More recently NLO corrections from electroweak
interactions \cite{Czakon:2017wor,Czakon:2017lgo} were also incorporated.
A more complete modelling of pair production including decay and off-shell
effects is available to NLO accuracy in QCD \cite{Bevilacqua:2010qb,
  Denner:2010jp} in the case of the di-lepton channel and more
recently also for the semi-leptonic channel
\cite{Denner:2017kzu}. These results were extended to pair
production in association with a jet \cite{Bevilacqua:2015qha,
  Bevilacqua:2016jfk}, which is of relevance for inclusive production
at NNLO.

Corrections to the top-quark decay process are known through NNLO in QCD
\cite{Brucherseifer:2013iv, Gao:2012ja}. This has allowed for a
partial prediction of the pair-production differential cross sections
with decay modelled within the Narrow Width Approximation (NWA)
\cite{Gao:2017goi}. The only missing piece of information is the exact
contribution from NNLO production followed by LO decay of the top
quarks. This requires the knowledge of polarised two-loop amplitudes
for this process, which is the subject of this publication.

The evaluation of the polarised two-loop amplitudes closely follows the lines of
\cite{Baernreuther:2013caa}. To obtain spin and color dependence of the
amplitudes, we use projection techniques, which were also successfully
applied in various two-loop calculations, for instance \cite{Gehrmann:2015ora}.
The most demanding part of this calculation is the
reduction and evaluation of involved scalar integrals.
The appearing scalar integrals can be reduced to the same set of master integrals
as those involved in the evaluation of the spin-summed amplitude.
The evaluation of these master integrals uses
a variety of analytical and numerical techniques. Exploiting the system of
differential equations obeyed by these master integrals is the core idea behind
these methods. Most of the physical phase space region can be accessed by
solving the differential equations numerically. The regions of phase space
that contain physical singularities cannot be reliably accessed using
numerical integration. We perform deep power-logarithmic expansions around
these singularities in order to obtain precise values for the master integrals.
We provide the results in terms of an expansion around the production threshold,
a high energy expansion, as well as an interpolation grid. To present and discuss
some features of our results, we recast the obtained coefficients with
respect to a basis in color and spin space, in terms the spin density
matrix of the top quarks alone. Although our results are obtained with
numerical methods, there is also progress in the analytic evaluation
of master integrals for this process \cite{Bonciani:2008az,
  Bonciani:2009nb, Bonciani:2010mn, vonManteuffel:2013uoa, Bonciani:2013ywa}.

This paper is organised in the following way. In the next section we define the
spin and color structures into which the amplitudes are decomposed. We
also discuss the projection method we used to obtain the
coefficients. Afterwards, we describe the methods used to obtain
numerical values for the master integrals 
in the physical phase space region as well as the
improvements we made considering the choice of the master integral basis. Next,
we present and discuss the results for the obtained coefficients. We close with
conclusions and outlook.

\section{Structure of the amplitude}

\subsection{Spin and color structures for virtual amplitudes}

The production of heavy quark pairs at hadron-hadron colliders involves 
two partonic QCD processes at lowest multiplicity
\begin{align}
  gg \to Q\bar{Q} \;\;\; \text{and} \;\;\; q\bar{q} \to Q\bar{Q}\;.
\end{align}
The momenta are assigned as follows
\begin{align} \label{momentaassignment}
 g,q(p_1) + g,\bar{q}(p_2) \to t (p_3) + \bar{t} (p_4) \; ,
\end{align}
with on-shell conditions
\begin{eqnarray}
p_1^2 = p_2^2 = 0 \;, \qquad p_3^2 = p_4^2 = m_t^2 \; .
\end{eqnarray}
We define the following kinematic invariants 
\begin{align}
 s \equiv \left(p_1 + p_2 \right)^2 \;, \qquad
 t \equiv m_t^2 - \left(p_1 - p_3 \right)^2 \;, \qquad
 u \equiv m_t^2 - \left(p_2 - p_3 \right)^2 \;,
\end{align}
where the relation $s-t-u = 0$ holds as a consequence of the 
aforementioned on-shell conditions and momentum conservation. 
These invariants are related to the scattering angle $\theta$ of the top quark  
(with respect to the beam axis in $p_1$ direction) and the top quark velocity 
$\beta$ as  
\begin{align}
  t = \frac{s}{2}\left(1-\beta\cos\theta\right) \; , \qquad u =
  \frac{s}{2}\left(1+\beta\cos\theta\right)\; , \qquad \text{with}\;\;
  \beta = \sqrt{1-4\frac{m_t^2}{s}} \; .
\end{align}
The bare scattering amplitude can be expanded in a perturbative series in 
$\as = g_s^2/4\pi$ and reads up to second order 
\begin{align} \label{eq:CSdecomform}
  \ket{\mathcal{M}_{g,q}(\as,m_t,\ep)} =
    4\pi\as\left[
               \ket{\mathcal{M}_{g,q}^{(0)}(m_t,\ep)}
              +\left(\frac{\as}{2\pi}\right)
               \ket{\mathcal{M}_{g,q}^{(1)}(m_t,\ep)}
              +\left(\frac{\as}{2\pi}\right)^2
               \ket{\mathcal{M}_{g,q}^{(2)}(m_t,\ep)}
           \right]\;.
\end{align}  
To facilitate the calculation of polarized virtual amplitudes, we decompose 
them in terms of color and spin (Lorentz) structures in the color $\otimes$ 
spin space of external particles. 

The color and spin decompositions of virtual amplitudes  
can be written as 
\begin{align} \label{eq:CSdecomform2}
\ket{\mathcal{M}_{g,q}^{(l)}(m_t,\ep)} = \sum_{i,j} c^{(l)}_{ij}(m_t,s,t,\ep)
\ket*{C^{g,q}_i} \otimes \ket*{S^{g,q}_j} \; ,
\end{align}
where $l =$ number of loops, and the $\ket*{C^{g,q}_i}, \ket*{S^{g,q}_j}$ on the
right-hand side 
represent, respectively, the chosen basis structures in the color $\otimes$ spin space.  
We denote the state of the external particles by $\ket{a,b,c,d}$ in color and 
$\ket{h_1,h_2,h_3,h_4}$ in spin space, where $a,b$ and $h_1,h_2$
concern the initial state, while $c,d$ and $h_3,h_4$ the final state,
in the same order as for the kinematics. 
With this notation, the color and spin basis structures of
Eq.~\eqref{eq:CSdecomform2} can be written in full generality in the
case of a gluon initial state as
\begin{align}
  \braket{a,b,c,d}{C^g_i} &= (C^g_i)^{ab}_{cd} \; , \nonumber\\
  \braket{h_1,h_2,h_3,h_4}{S_i^g} &= \ep_1(h_1)_\mu \ep_2(h_2)_\nu
                  \bar{u}_3(h_3)(S_i)^{g\mu\nu}v_4(h_4) \; ,
\end{align}
and similarly for a quark initial state
\begin{align}
\braket{a,b,c,d}{C^q_i} &= (C^q_i)_{abcd} \; , \nonumber\\
\braket{h_1,h_2,h_3,h_4}{S^q_i} &= \bar{v}_2(h_2) \Gamma_i u_1(h_1)
                                  \bar{u}_3(h_3) \Gamma_i' v_4(h_4) \; .
\end{align} 
Notice that in our work, we do not choose any specific representation
of spinors or polarisation vectors. Thus, for example, $h_{3,4}$ are not
necessarily helicities in the case of the heavy quarks. By providing
results in terms of spin structures $S_i$, we allow to translate the amplitudes
to any particular polarisation basis. For phenomenological
applications, we provide the spin density matrix, which contains all
the necessary information in terms of spin vectors of the heavy
quarks, see Section~\ref{sec:spindensity}.

The color decomposition basis of amplitudes can be chosen straightforwardly.
For the $gg \rightarrow t\bar{t}$, we use the natural basis
\begin{align} \label{eq:ggbasis}
  C^g_1 &= ( T^a T^b)_{cd}\;,\nonumber\\
  C^g_2 &= ( T^b T^a)_{cd}\;,\nonumber\\
  C^g_3 &= \Tr\{ T^a T^b\}\delta_{cd}\;. 
\end{align}
In the case of quark annihilation in the initial state, 
the color basis reads
\begin{align}
  C^q_1 &= \delta_{ab}\delta_{cd} \; , \nonumber\\
  C^q_2 &= \delta_{ad}\delta_{cb}\;.
\end{align}

For each of the color structures $C_i$, we decompose the amplitudes
further in terms of spin (Lorentz) structures.  
To this end, we assume that all four external particles are confined
to $4$-dimensional space and are on-shell with physical polarization
states (i.e. $4$-dimensional equations of motion are satisfied).
Under this condition, we have in total $2^4 = 16$ different
physical helicity configurations both in the $gg \rightarrow t\bar{t}$
and the $q\bar{q} \rightarrow t\bar{t}$ process. 
Additional symmetry properties enjoyed by the amplitudes can 
lead to relations which further reduce the number of 
linearly independent structures.
Indeed, on top of the aforementioned kinematic constraints, 
QCD interactions are invariant with respect to parity,
under which the helicity of each of the four external particles 
is flipped, while the color structures are left unchanged. 
This symmetry then reduces the linearly independent spin structures
down to 8 (in $4$-dimensional space), both in the
$gg \rightarrow t\bar{t}$ and $q\bar{q} \rightarrow t\bar{t}$ cases. 
QCD interactions are also invariant under charge conjugation. However,
this operation also involves the color structure. For this reason, we
did not impose C-symmetry when   determining the basis of Lorentz
structures for the color-stripped amplitudes. For the same reason,
implications from Bose-symmetry between 
the two gluons are not considered at this point, but rather used as a
test at the end of the calculation. 

At this point, we discuss the particulars for the two specific
amplitudes, since additional symmetry properties are 
process dependent.

Let us first consider the $gg \rightarrow t\bar{t}$ case. 
Here, we assume that both polarisation vectors are orthogonal to both
of the initial state momenta, $p_1$ and $p_2$ (see
Eq.~\eqref{momentaassignment}). This reduces the number of degrees of
freedom to the physical two. Spin sums should, therefore, be performed with
\begin{align}
  \sum_\text{h} \ep^{*}_{\mu}(h) \ep_{\nu}(h) = 
    \left(
      -g_{\mu\nu}+\frac{p_{1\mu}p_{2\nu}+p_{1\nu}p_{2\mu}}{p_1\cdot
  p_2} \right) \; .
\end{align}
After stripping off the external wave functions we choose the following
set of 8 Lorentz structures, with suppressed spinor indices
\begin{align}
S_1^{g \mu \nu} &= \frac{1}{s}\left(\g^{\mu} p_3^{\nu}+\g^{\nu} p_3^{\mu}\right) \; ,
& S_2^{g \mu \nu}  = \frac{m_t}{s} \, g^{\mu \nu} \mathbf{1} \; ,
\nonumber \\[0.2cm]
S_3^{g \mu \nu} &= \frac{1}{s \, m_t} \, p_3^{\mu} \, p_3^{\nu} \mathbf{1} \; ,
& S_4^{g \mu \nu}  = \frac{1}{s \, m_t^2} \, \slashed{p}_1 p_3^{\mu} p_3^{\nu} \; ,
\nonumber \\[0.2cm]
S_5^{g \mu \nu} &= \frac{1}{s} \, \slashed{p}_1 g^{\mu \nu} \; ,
& S_6^{g \mu \nu}  = \frac{1}{s \, m_t} \, \slashed{p}_1\left(\g^{\nu}
  p_3^{\mu} +\g^{\mu} p_3^{\nu} \right) \; ,
\nonumber\\[0.2cm]
S_7^{g \mu \nu} &= \frac{1}{s} \left(\g^{\mu} p_3^{\nu}-\g^{\nu} p_3^{\mu}\right) \; ,
& S_8^{g \mu \nu}  = \frac{m_t}{s} \left(\slashed{p}_1 g^{\mu
  \nu}-\slashed{p}_1 \g^{\mu}\g^{\nu}\right) \; .
\end{align}
The additional factors of $m_t$ and $s$ are inserted such that all
structures are dimensionless once multiplied with spinors (which are
assumed to have mass dimension 1/2). The structures are grouped according to
whether they are symmetric ($S_1$ to $S_6$) or anti-symmetric
($S_7$ and $S_8$) under the exchange of $\mu \leftrightarrow \nu$.  It
can be checked explicitly that each of the above Lorentz structures is
mapped back to itself under the parity transformation (up to a
phase factor). The Gram determinant of this set of structures is not
identically zero, assuring that they are linearly independent.

In case of the $q\bar{q} \rightarrow t\bar{t}$ process with massless
initial-state quarks and limited to QCD interactions, the massless quark line is 
disconnected from the massive top-quark line. 
Chirality conservation in QCD, therefore, implies that only half of 
the helicity configurations of the initial-state massless quarks are non-zero. Once this additional constraint is accounted on top of those aforementioned 
ones, one finds that there are only four independent helicity
amplitudes left.
We therefore choose the following set of 4 Lorentz structures 
of the form $S = \Gamma \otimes \Gamma'$ 
($\Gamma$ denotes a string of $\gamma$ matrices)
\begin{equation}
S^q_1 = \frac{1}{s \, m_t} \, \slashed{p}_3 \otimes \mathbf{1} \;, \quad
S^q_2 = \frac{1}{s \, m_t^2} \, \slashed{p}_3 \otimes \slashed{p}_1 \;, \quad
S^q_3 = \frac{1}{s} \, \g^{\mu} \otimes \g_{\mu} \;, \quad
S^q_4 = \frac{1}{s \, m_t} \, \g^{\mu} \otimes ( \slashed{p}_1\g_{\mu}
) \;,
\end{equation}
where the left-hand side of the $\otimes$ symbol concerns the massless
fermion line, while the right-hand side concerns the massive fermion line. 
Again, the linear independence of these structures can be verified via
the Gram determinant.

The coefficient functions of the above color and spin decomposition of
amplitudes
can be extracted by performing the usual projection procedure. In short, 
the projection of the virtual amplitude onto each of the chosen basis 
structures gives an equation linear in the coefficient functions.   
The collection of all projections onto the linearly independent, complete
set of basis structures then forms an invertible linear algebraic equation
system in the coefficient functions, which can be solved straightforwardly. 
The coefficient matrix of this linear algebraic equation system is identical
to the Gram matrix of the chosen basis.

We note already at this point that in our calculation, the structures
$S^g_6$ and $S^q_4$ have vanishing coefficients for all color structures.

Even though we performed our calculations with the basis specified in
Eq.~\eqref{eq:ggbasis}, it is possible to express the coefficient
functions for the gluon channel in terms of the orthonormal basis
\begin{align}
  C_\mathbf{8_S}^g &= \sqrt{\frac{2N_C}{(N_C^2-1)(N_C^2-4)}} \, \Big(
  C_1^g+C_2^g-\frac{2}{N_C} C_3^g \Big) \; , \\[0.2cm]
  C_\mathbf{8_A}^g &= \sqrt{\frac{2}{N_C(N_C-1)}} \, \big(
  C_1^g-C_2^g \big) \; ,\\[0.2cm]
  C_\mathbf{1}^g   &= \frac{2}{\sqrt{N_C(N_C-1)}} \, C_3^g \; ,
\end{align}
where $N_C = 3$ is the number of colors, and $\mathbf{8_S}$,
$\mathbf{8_A}$ denote the symmetric and anti-symmetric octet states
respectively, while $\mathbf{1}$ the singlet state.  The advantage of
this choice of basis is that there is no mixing between the color
structures when calculating color summed amplitudes. On the other
hand, with these structures the coefficient functions exhibit a simple
Bose symmetry. Indeed, for the spin structures
$S_1^{g\,\mu\nu},S_2^{g\,\mu\nu},S_3^{g\,\mu\nu},S_7^{g\,\mu\nu},S_8^{g\,\mu\nu}$
the coefficients of $C_\mathbf{8_S}^g$ and $C_\mathbf{1}^g$ are
symmetric under the exchange $\cos \theta \to -\cos \theta$, while the
coefficient of $C_\mathbf{8_A}^g$ is anti-symmetric under this
transformation. For the spin structures $S_4^{g\,\mu\nu}$ and
$S_5^{g\,\mu\nu}$ the situation is reversed, the coefficients of
$C_\mathbf{8_S}^g$ and $C_\mathbf{1}^g$ are anti-symmetric while
$C_\mathbf{8_A}^g$ have symmetric coefficients. These properties are
consistent with the numerical results, and constitute a test of the
calculation.

\subsection{Ultraviolet and infrared renormalisation}

The chosen color $\otimes$ spin basis may be used in $d$-dimensions
after extension of the spin structures by evanescent combinations. For
physical applications, however, we should only need 4-dimensional
quantities. Due to the presence of infrared singularities, meaningful
amplitudes are only obtained after the usual ultraviolet
renormalisation followed by infrared subtraction (multiplicative
renormalisation). This procedure results in so-called finite
remainders, which are, however, scheme dependent.

The UV renormalized amplitude reads 
\begin{align}
  \ket{\mathcal{M}_{g,q}^R(\asnf,m,\mu,\ep)} =
    \left(\frac{\mu^2e^{\gamma_E}}{4\pi}\right)^{-2\ep}
    Z_{g,q}Z_Q\ket{\mathcal{M}_{g,q}^0(\asb,m^0,\ep)} \; ,
\end{align}
where we used the on-shell wave function renormalisation constants
$Z_g$, $Z_q$ and $Z_Q$. The renormalised heavy quark mass $m$ is
related to the bare mass by $m^0=Z_m m$. The coupling constant is
renormalized in the $\MSbar$ scheme with $\nf = \nl + \nh$
active flavours
\begin{align}
  \asb = \left(\frac{e^{\g_E}}{4\pi}\right)^{\ep}\mu^{2\ep}
  Z_{\as}^{(\nf)} \asnf(\mu) \; .
\end{align}
As argued in \cite{Baernreuther:2013caa} a decoupling of the heavy flavours
from the running of $\as$ is necessary to correctly accommodate for
heavy quark mass effects in regimes where the produced heavy quarks
are not very relativistic. This decoupling can be achieved by the replacement
\begin{align}
  \asnf = \zeta_{\as}\asnl \; ,
\end{align}
where $\zeta_{\as}$ is the decoupling constant.

The wave-function and the coupling renormalisation (including decoupling) act
multiplicatively on the amplitudes and, therefore, also on the coefficients
$c_{ij}$. The mass renormalization counter term, on
the other hand, requires an additional decomposition of the lower
order amplitudes into color $\otimes$ spin structures.
The necessary renormalisation and decoupling constants are given in
the appendix  \ref{sec:RConst}.

The UV renormalized coefficient functions still contain infrared
divergences. However, the infrared structure is known in terms of
lower order amplitudes \cite{Aybat:2006mz,Mitov:2009sv,Becher:2009qa,
      Becher:2009kw,Czakon:2009zw,Ferroglia:2009ii,
       Mitov:2010xw}, and can be extracted from the UV renormalised
     amplitude
\begin{align}
\ket{\mathcal{M}_n^{(0)}} &= \ket{\mathcal{F}_n^{(0)}}\;\;,\\
\ket{\mathcal{M}_n^{(1)}} &= \mathbf{Z}^{(1)}\ket{\mathcal{M}_n^{(0)}}
                             +\ket{\mathcal{F}_n^{(1)}}\;\;,\\
\ket{\mathcal{M}_n^{(2)}} &= \mathbf{Z}^{(2)}\ket{\mathcal{M}_n^{(0)}}+
                             \mathbf{Z}^{(1)}\ket{\mathcal{F}_n^{(1)}}+ 
                             \ket{\mathcal{F}_n^{(2)}}\\
                          &= \left(\mathbf{Z}^{(2)}-
                                   \mathbf{Z}^{(1)}\mathbf{Z}^{(1)}
                             \right)\ket{\mathcal{M}_n^{(0)}}
                             + \mathbf{Z}^{(1)}\ket{\mathcal{M}_n^{(1)}}
                             +\ket{\mathcal{F}_n^{(2)}} \;, 
\end{align}
where $\ket{\mathcal{F}_n}$ is the finite remainder amplitude, we are
interested in. $\mathbf{Z} =
\mathbf{1}+\mathbf{Z}^{(1)}+\mathbf{Z}^{(2)}+\order{\as^3}$ 
is the IR renormalization constant. It is an operator in
color space and can be obtained from its renormalisation group equation
\begin{align}
  \frac{\text{d}}{\text{d}\ln \mu_R}
     \mathbf{Z}\left(\ep,\{p_i\},\{m_i\},\mu_R\right)
       = -\mathbf{\Gamma}\left(\{p_i\},\{m_i\},\mu_R\right)
          \mathbf{Z}\left(\ep,\{p_i\},\{m_i\},\mu_R\right) \; ,
\end{align}
where the anomalous dimension $\mathbf{\Gamma}$ is given in the
appendix \ref{sec:RConst}. 
Since the $\mathbf{Z}$ operator acts in color-space, the terms 
$\mathbf{Z}^{(i)}\ket{\mathcal{M}_n^{(k)}}$ have to be projected back onto the
color structures to obtain the corresponding counter terms for the
coefficients. The minimal definition of the renormalisation operator
$\mathbf{Z}$, which consists of poles in the dimensional
regularisation parameter only, specifies our IR renormalisation scheme
uniquely.

In \cite{Czakon:2013hxa}, it was shown that the triple-color
correlators of the soft anomalous dimension matrix cannot contribute to spin
and color summed matrix elements. Since we keep
color and spin dependence this is no longer true in our case. In fact,
our calculation is the first to rely on the coefficient
\begin{align*}
\sum\limits_{(I,J)}\sum\limits_k\,i\,f^{abc}\,
{\bf T}_I^a\,{\bf T}_J^b\,{\bf T}_k^c\,f_2\left(\beta_{IJ},
\ln\frac{-\sigma_{Jk}\,v_J\cdot p_k}{-\sigma_{Ik}\,
v_I\cdot p_k}\right) \; ,
\end{align*}
to correctly obtain all poles of the coefficients functions. In consequence, it
constitutes the first non-trivial cross-check of this contribution to
the soft anomalous dimension matrix,
which was originally derived in \cite{Ferroglia:2009ii}.

\subsection{Spin density matrix}
\label{sec:spindensity}
For illustration of our results, we choose to recast the
amplitude into a more convenient form. In general, since each
coefficient has a real and imaginary part, our calculation yields 54 real
functions. However, not all of them enter independently into physical
predictions. Therefore, we also evaluate the spin
density matrix, which contains all the necessary information
on the top-quark spin dependence and is sufficient for
phenomenological applications.

The spins of the top-quarks in their rest frame can be
described by two normalised spin 3-vectors
$\hat{\mathbf{s}}_{t}$ and $\hat{\mathbf{s}}_{\bar{t}}$.
They correspond to two four-vectors $s_t$ and $s_{\bar{t}}$ in the center of
mass frame which have the properties
\begin{align}
 s_{t}^2 = s_{\bar{t}}^2 = -1 \;\;\; \text{and}\;\;\; p_3 \cdot s_t = p_4 \cdot
s_{\bar{t}} = 0 \;.
\end{align}
The vectors $s_t$ and $s_{\bar{t}}$ enter the matrix element through the
insertion of the spin projectors
\begin{align}
  u(p_3, s_t)\bar{u}(p_3, s_t) &=
\left(\slashed{p}_3+m\right)\frac{1}{2}\left(1+\g_5\slashed{s}_t\right)
  \; , \\
  v(p_4, s_{\bar{t}})\bar{v}(p_4, s_{\bar{t}})  &=
\left(\slashed{p}_4-m\right)\frac{1}{2}\left(1+\g_5\slashed{s}_{\bar{t}}\right)\;.
\end{align}
Since we work with finite remainders without any divergences, the
presence of the $\gamma_5$ matrix does not constitute any
complication. Indeed, the spin density matrix is simply evaluated in 4
dimensions.
The two-loop contribution to the spin-density matrix for both partonic
processes can be decomposed as
\begin{align}
\mathcal{R}^{\rm 2-loop}_{q,g}(s_{t},s_{\bar{t}})
   &= \ 2\Re\braket{\mathcal{M}^{0}_{q,g}}{\mathcal{M}^{2}_{q,g}}(s_t,s_{\bar{t}}) 
   = A_{q,g} +
    \left(C \right)_{q,g}
         \biggl(
           (s_t \cdot s_{\bar{t}})
         \biggr) +\nonumber\\  
    &\left(B_t\right)_{q,g}
         \biggl(
           \ep^{\mu\nu\alpha\beta}p_{1\mu}p_{2\nu}p_{3\alpha}s_{t\beta}
         \biggr) + 
    \left(B_{\bar{t}}\right)_{q,g}
         \biggl(
           \ep^{\mu\nu\alpha\beta}p_{1\mu}p_{2\nu}p_{3\alpha}s_{\bar{t}\beta}
         \biggr) +\nonumber\\
    &\left(D_1\right)_{q,g}
         \biggl((p_1\cdot s_t)(p_1\cdot s_{\bar{t}})\biggr)+
    \left(D_2\right)_{q,g}
         \biggl((p_2\cdot s_t)(p_2\cdot s_{\bar{t}})\biggr)+\nonumber\\
    &\left(E_{12}\right)_{q,g}
         \biggl((p_1\cdot s_t)(p_2\cdot s_{\bar{t}})\biggr)+
    \left(E_{21}\right)_{q,g}
         \biggl((p_2\cdot s_t)(p_1\cdot s_{\bar{t}})\biggr) \; .
\end{align}
These functions are related to the 2-loop components of the spin
density matrix $R^{q,g}$ as defined in
\cite{Bernreuther:1993hq} through
\begin{align}
\mathcal{R}_{q,g}^{\rm 2-loop} = \frac{1}{4} \Tr
\left[ 
R^{q,g}(\mathbf{1}+\hat{\mathbf{s}}_t\sigma)\otimes(\mathbf{1}+\hat{\mathbf{s}}_{\bar{t}}\sigma)
\right]\biggr|_{\rm 2-loop} \; .
\end{align}

The coefficients of the occurring structures are functions of $\cos
\theta$ and $\beta$ only. In pure QCD C,P and CP invariance hold and
imply that $B_t = B_{\bar{t}}=B$ as well as $D_1 = D_2 = D$
\cite{Bernreuther:1993hq} for both channels.  Therefore we are left with
\begin{align}
 \mathcal{R}_{q,g}^{\rm 2-loop}&
   =
    A_{q,g} +
    \left(B\right)_{q,g}
         \biggl(
           \ep^{\mu\nu\alpha\beta}p_{1\mu}p_{2\nu}p_{3\alpha}s_{t\beta}
           +\ep^{\mu\nu\alpha\beta}p_{1\mu}p_{2\nu}p_{3\alpha}s_{\bar{t}\beta}
         \biggr) \nonumber\\
    &+\left(C\right)_{q,g}
         \biggl(
          (s_t\cdot s_{\bar{t}}) 
         \biggr)+
     \left(D\right)_{q,g}
         \biggl(
                (p_1\cdot s_t)(p_1\cdot s_{\bar{t}})+
                (p_2\cdot s_{t})(p_2\cdot s_{\bar{t}})
         \biggr)\nonumber\\
    &+\left(E_{12}\right)_{q,g}
         \biggl((p_1\cdot s_t)(p_2\cdot s_{\bar{t}})\biggr)+
    \left(E_{21}\right)_{q,g}
         \biggl((p_2\cdot s_t)(p_1\cdot s_{\bar{t}})\biggr) \; .
\label{eq:sdm-final}
\end{align}
In the gluon case we have an additional bose-symmetry which implies that the
functions $A_g,C_g,D_g$ are symmetric in $\cos \theta$ and that $B_g$ has to be
an antisymmetric function in $\cos \theta$. It also implies the relation
$E_{12 g}(\cos\theta) = E_{21 g}(-\cos\theta)$. 

\section{Scalar integrals}
\label{sec:methods}
The coefficients functions $c_{ij}$ are given by linear combinations
of a large number of scalar integrals with rational coefficients in 
$s,t,m^2$ and $\ep$. 
These scalar integrals are expressed through linear combinations of 
master integrals using an Integration-by-Parts (IBP) reduction. 
We can rewrite the coefficients, as well as the master integrals, in terms
of dimensionless variables $m_s= m^2/s$ and $x = t/s$. From the IBP relations we
can obtain a system of differential equations for the master integrals
\begin{align}
  m_s \frac{\partial}{\partial m_s} \vec{I}(m_s,x,\ep) &=
    A^{(m_s)}(m_s,x,\ep)\vec{I}(m_s,x,\ep) \; , \\
  x \frac{\partial}{\partial x} \vec{I}(m_s,x,\ep) &=
    A^{(x)}(m_s,x,\ep)\vec{I}(m_s,x,\ep) \; ,
\end{align}
where $A^{(m_s)}$ and $A^{(x)}$ are matrices whose elements are rational
functions in $m_s,x$ and $\ep$.

\begin{figure}
  \centering
  \includegraphics[width=5cm]{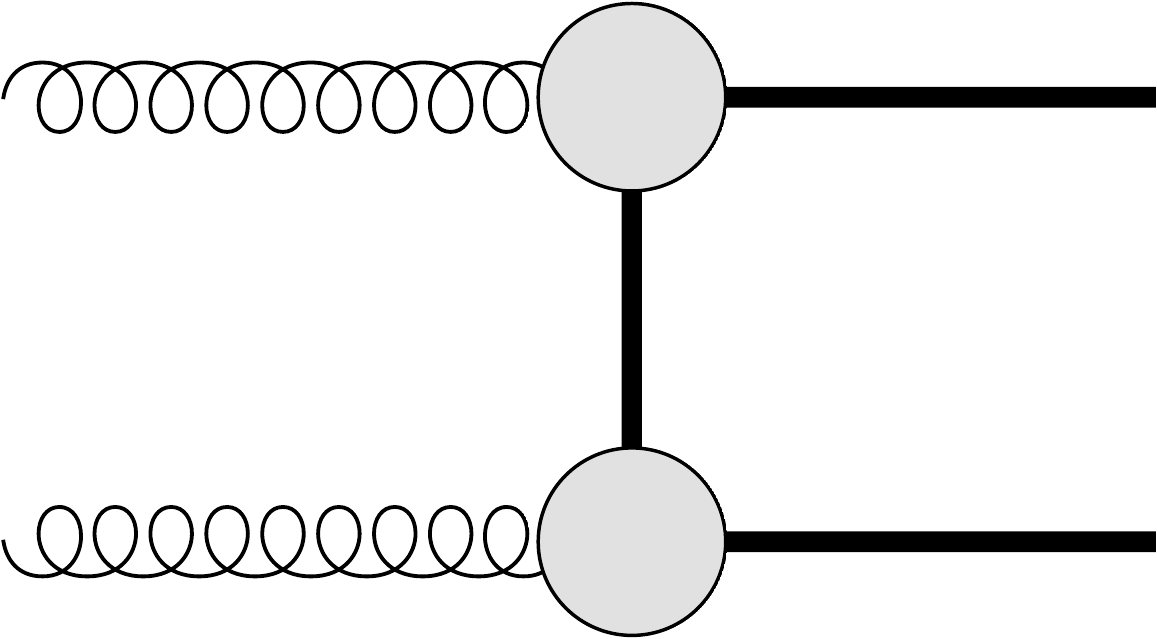}
  \caption{Class of diagrams leading to enhanced matrix elements at high energy 
           and low scattering angle.}
  \label{fig:tchannel}
\end{figure}

We do not choose the same set of master integrals as in the spin summed
calculation. There is an enhancement of the matrix elements
at high energies and small/large scattering angles resulting from diagrams
of a $t$/$u$-channel type as indicated in Fig.~\ref{fig:tchannel}. These
enhancements require numerically very stable results for the master
integrals in these phase space regions. For this reason, we decided to
try a basis for a subset of the integrals, which corresponds to the
$\epsilon$-form of the differential equation (see next section). Our
hope was that in this basis the numerical evaluation will become more
stable. Ultimately, this turned out not to be the case. We stress,
nevertheless, that the results obtained in the old and new bases for
the spin summed amplitudes agree to several digits, within the
accuracy of the calculation. 
 
\subsection{Canonicalization}

With the hope to achieve a better stability when numerically solving the 
differential equations for master integrals involved in the 
two-loop $gg \rightarrow t \bar{t}$ process, we choose to put 
the equation system partially into the $\epsilon$-form \cite{Henn:2013pwa},
where the right-hand side of the differential equation
system is proportional to $\epsilon = \frac{d-4}{2}$ 
and the singularities are only simple poles
in the kinematic variables. 
Algorithmic approaches have been devised to arrive at the 
$\epsilon$-form for a given differential equation system for master-integrals 
in a single variable \cite{Lee:2014ioa,Lee:2017oca}. They have been
implemented in  \textsc{Fushsia} \cite{Gituliar:2017vzm} and
\textsc{Epsilon} \cite{Prausa:2017ltv}  which are publicly
available. In the case of multiple variables there also exists an
algorithm, presented in Refs.~\cite{Meyer:2016slj, Meyer:2017joq} and
implemented in a program called \textsc{Canonica}.
It is well known that, for a given set of master integrals,
an $\epsilon$-form is not always achievable by a rational
transformation of the integral basis. It is also not always possible
\cite{Henn:2014qga, Adams:2017tga, Lee:2017oca} even with more general
transformations.
In particular, Ref.~\cite{Lee:2017oca} provides a strict criterion 
for the existence of an $\epsilon$-form in the case of master integrals of 
a single variable. In particular, the 4-dimensional homogeneous part
of the differential equation system typically corresponds to
high-order Picard-Fuchs differential equations that do not factorize  
completely \cite{Adams:2017tga}.
The simplest counter example is given by the differential equations of 
the master integrals of the two-loop sunset diagrams with identical masses 
\cite{Caffo:1998du, Laporta:2004rb}, where solutions involve elliptic integrals.  

The topology of the two-loop sunset diagrams with equal masses appears in the 
IBP-reduction of the master integrals of the two-loop $gg \rightarrow
t\bar{t}$ diagrams. 
Thus, it is not a surprise that the full system of differential equations of
the 422 master integrals involved cannot be completely put into 
the $\epsilon$-form. In addition, a considerable amount of sectors require
individual coordinate transformations in order to arrive at their respective 
$\epsilon$-forms by rational transformations (in the new variables). 
Since we would like to numerically integrate the full differential equation system
of all master integrals in one go, we are then forced to divide the 
master integrals into two subsets: those that can be directly put into 
the $\epsilon$-form via a rational transformation in the original variables
and those that cannot. The second subset essentially consists of master integrals
fulfilling any of the following three conditions:
\begin{itemize} 
\item [1)] 
their expressions involve elliptic integrals;
\item [2)] 
coordinate transformations are required in order to reach their $\epsilon$-form; 
\item [3)]
their derivatives involve any one of the aforementioned two kinds of master integrals. 
\end{itemize}
Under such tight selection criteria, there are only 65 master integrals
that can be directly transformed into the basis observing the $\epsilon$-form
(in the original variables). They are identified and then subsequently 
moved to the front of the differential equation system of the 422
master integrals,
without spoiling the block-wise triangular structure of the
differential equation system.  
The numerical evaluation of the complicated master integrals
are expected to benefit from 
the $\epsilon$-form of these 65 master integrals.
The differential equation system of these 65 master integrals in question 
involves more than one variable\footnote{We treat the triangle graphs as part
of box topologies. In principle, one could single out this class of
graphs and solve them separately. In this case, there is a basis in
which they only depend on one variable. We were interested in trying
canonicalization in the multivariate case.}, and we employ the package
\textsc{CANONICA} \cite{Meyer:2017joq} to find the rational
transformation matrix  
needed for obtaining the $\epsilon$-form. 
A few modifications of the program were made in order to tackle this 
65-by-65 system with less time consumption.

As a side remark, we would like to briefly mention the following point.  
Due to the existence of \textit{remnant} rational transformations that preserve 
the $\epsilon$-form of a differential equation system, the new basis integrals 
defined by the rational transformation matrix returned by the package 
\textsc{CANONICA} \cite{Meyer:2017joq} are not guaranteed to be of
uniform weight \cite{Henn:2013pwa,Henn:2014qga}. 
Upon a closer examination, we find that in general not all of these remnant 
rational transformations respecting the $\epsilon$-form (if it exists) 
are of weight 0 according to the counting rules laid in 
\cite{Henn:2013pwa,Henn:2014qga}.

To be more specific about this, we find this remnant freedom to be the following. 
Under any rational transformation mixed with any coordinate-transformation
under the condition of keeping the resulting differential system still rational 
in the new variables, the \textit{remnant} rational transformations 
that preserve the $\epsilon$-form of the differential equation system 
(assuming it exists and is represented by $\epsilon \textup{d}\tilde{A}$), 
read
\begin{equation}
\hat{T}_{R}(\epsilon) = \hat{T}_{I}(\epsilon) \hat{C} \; ,
\end{equation}
where $\hat{C}$ can be any invertible constant matrix of rational numbers 
of the same dimension as the $\epsilon$-form coefficient-matrix 
$\textup{d}\tilde{A}$. $\hat{T}_{I}(\epsilon)$ is independent
of kinematics but possibly possesses some non-trivial
$\epsilon$-dependence. It can be any element from the invariance symmetry group of the coefficient-matrix 
$\textup{d}\tilde{A}$ with matrix elements being Laurent-polynomials in 
$\epsilon$ with rational numerical coefficients. 
In other words, $\hat{T}_{I}(\epsilon)$ is a matrix that is invertible and commutes with 
$\textup{d}\tilde{A}$ with all its matrix-elements living in the ring of 
$\epsilon$ over rational numbers.
Owing to its invertibility and the fact that it consists of rational
numbers only, the 
matrix $\hat{C}$ always preserves the uniform weight feature
of the vector on which it acts (if this feature is there in the first place). 
However, some of $\hat{T}_{I}(\epsilon)$ with non-trivial $\epsilon$-dependence 
can turn a list of uniform weight master-integrals into a list of non-uniform-weight 
integrals, and vice versa, even though both perfectly observe $\epsilon$-form 
differential equations. 
Since there is no reference to the concrete boundary conditions of 
the differential equations in the process of finding the rational 
transformation done by the package \textsc{CANONICA}, it is therefore not 
guaranteed that the solutions of the $\epsilon$-form differential 
equations thus obtained are of uniform weight.

\subsection{Master integral evaluation}

We subdivide the physical phase space region into three regions, the high energy
limit where $m_s \to 0$, the threshold region $\beta \to 0$, and the
``bulk'' which describes the rest of the phase space region.

For the numerical integration of the new set of master integrals, high precision
boundaries are needed. They are obtained from the power-logarithmic expansion
in the high energy limit from the original set of master integrals.
The first few terms of those expansions were obtained with Mellin-Barns
techniques using the MB package \cite{Czakon:2005rk}.
These expansions are exact in $t$, where in
some cases the differential equations were used to get the exact behaviour from
the limit $t=0$. In this double limit $m^2\to0$ and $t\to0$ the integrals were
evaluated numerically with very high precision and then resummed with PSLQ
algorithm \cite{pslq} or XSummer \cite{Moch:2005uc}.
These first terms were
used to derive deep expansions in $m_s$ by using the available
differential equations. These deep expansions were subsequently used to
compute high precision boundaries for the numerical integration. 

Starting from the numerical results obtained from the deep power logarithmic
expansion, we perform a numerical integration along contours in the complex
plane. In our programs we incorporate software from \cite{vode} for solving the
differential equations and \cite{qd} to handle higher precision numbers.
The endpoints of the contours define an interpolation grid. In the
region which is accessible with this method, i.e. where the coefficient functions 
are not too singular, we evaluate the amplitudes using this interpolation grid. 
The sampling points are the same as in
\cite{Czakon:2013goa,Baernreuther:2012ws,Czakon:2012zr,Czakon:2012pz} with an
extension to higher values of $\beta$. 

In the limit $\beta \to 0$ some master integrals show singular behavior and are
difficult to obtain with the method of numerical integration.
We perform a deep power-logarithmic expansion of the master
integrals in $\beta$ by again exploiting the differential equations up to
$\order{\beta^{50}}$ and $\order{\ln^{10}\beta}$.
This expansion is done for several fixed angles $\cos \theta$ with unknown
boundary conditions, which are finally determined by matching to the results
obtained by numerical integration.

\section{Results}

\begin{figure}[th!]
\subfigure[]{
               \includegraphics[width=7.5cm]{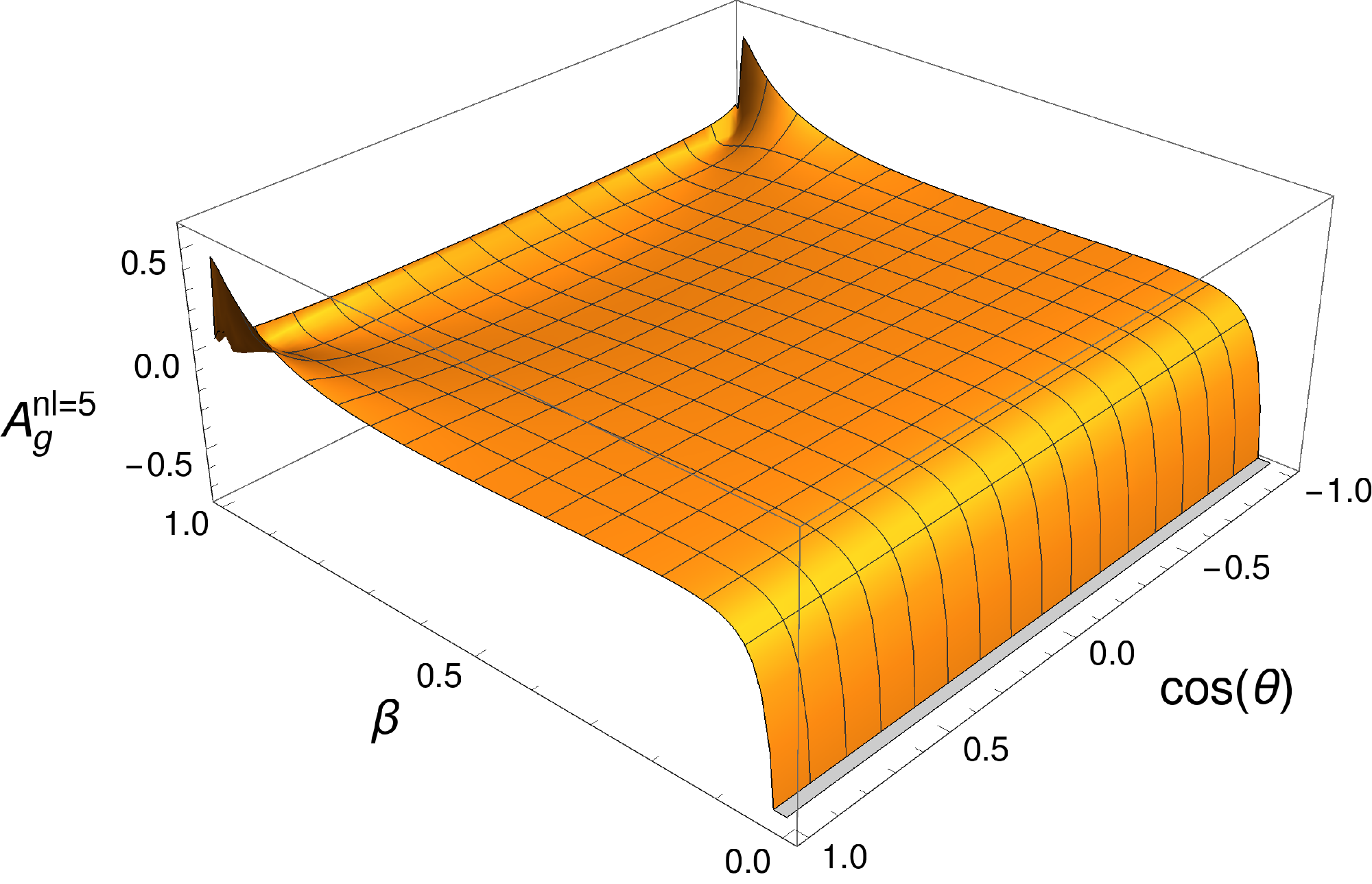}
              }%
\subfigure[]{
               \includegraphics[width=7.5cm]{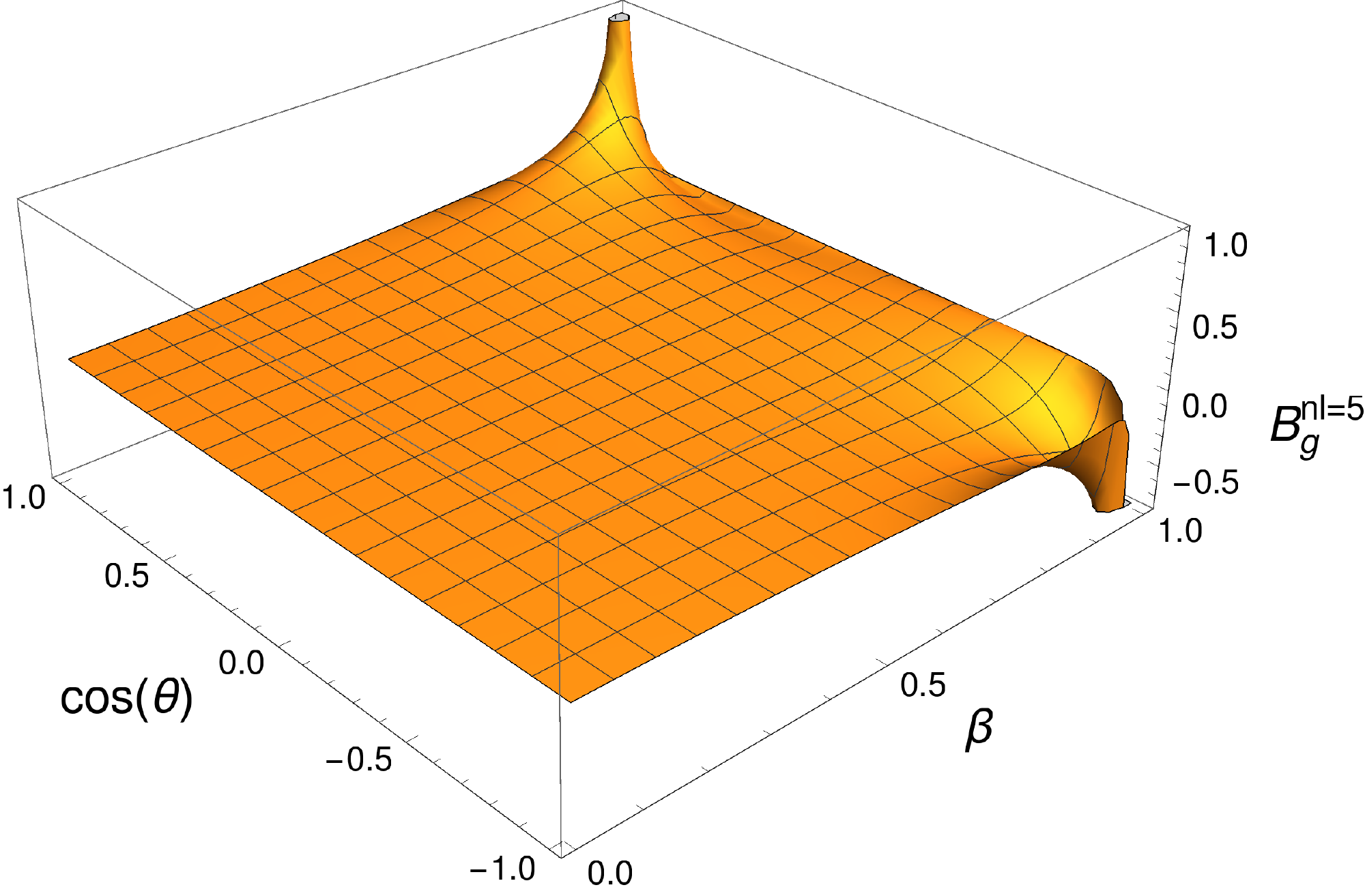}
              }%

\subfigure[]{
               \includegraphics[width=7.5cm]{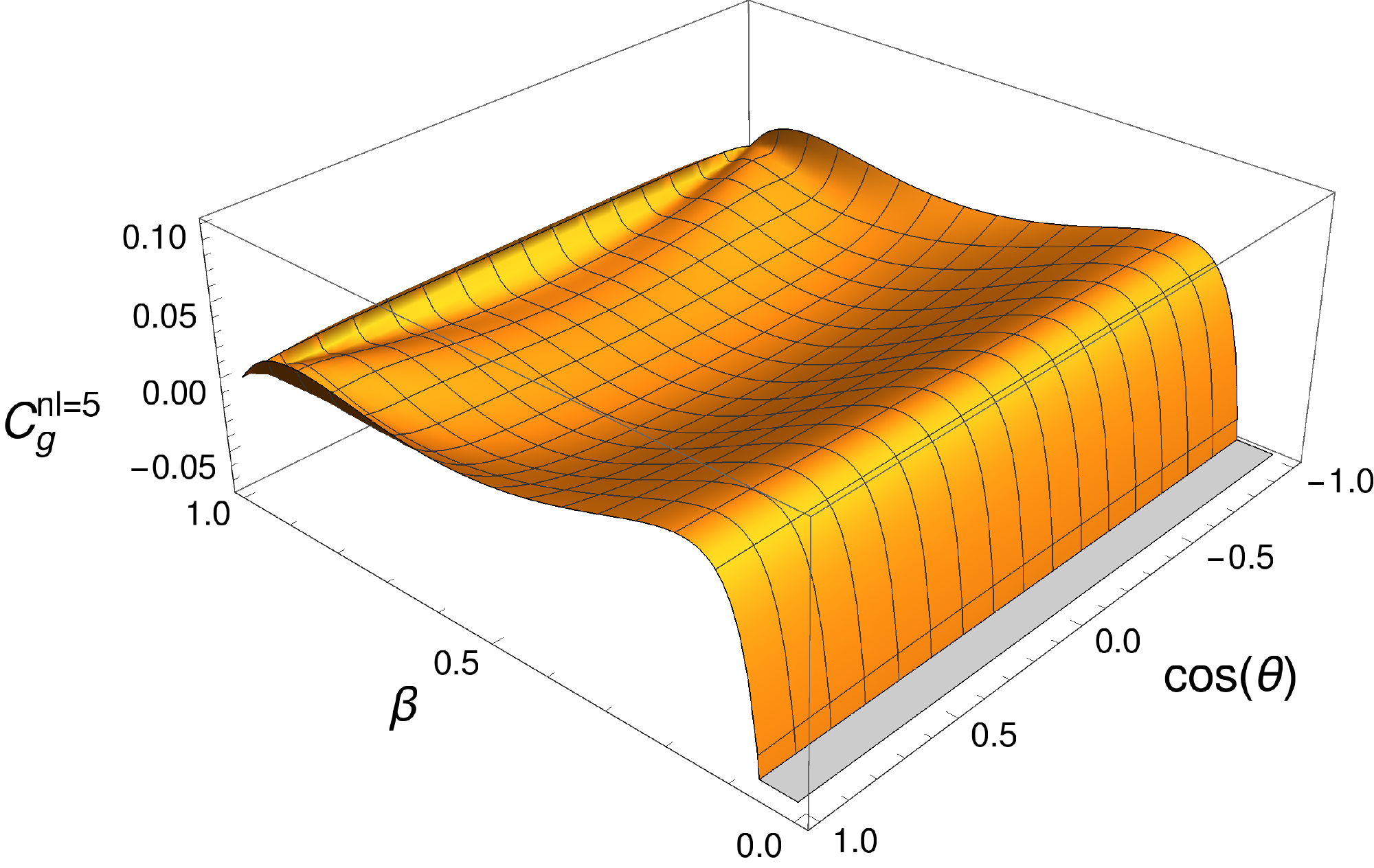}
              }%
\subfigure[]{
               \includegraphics[width=7.5cm]{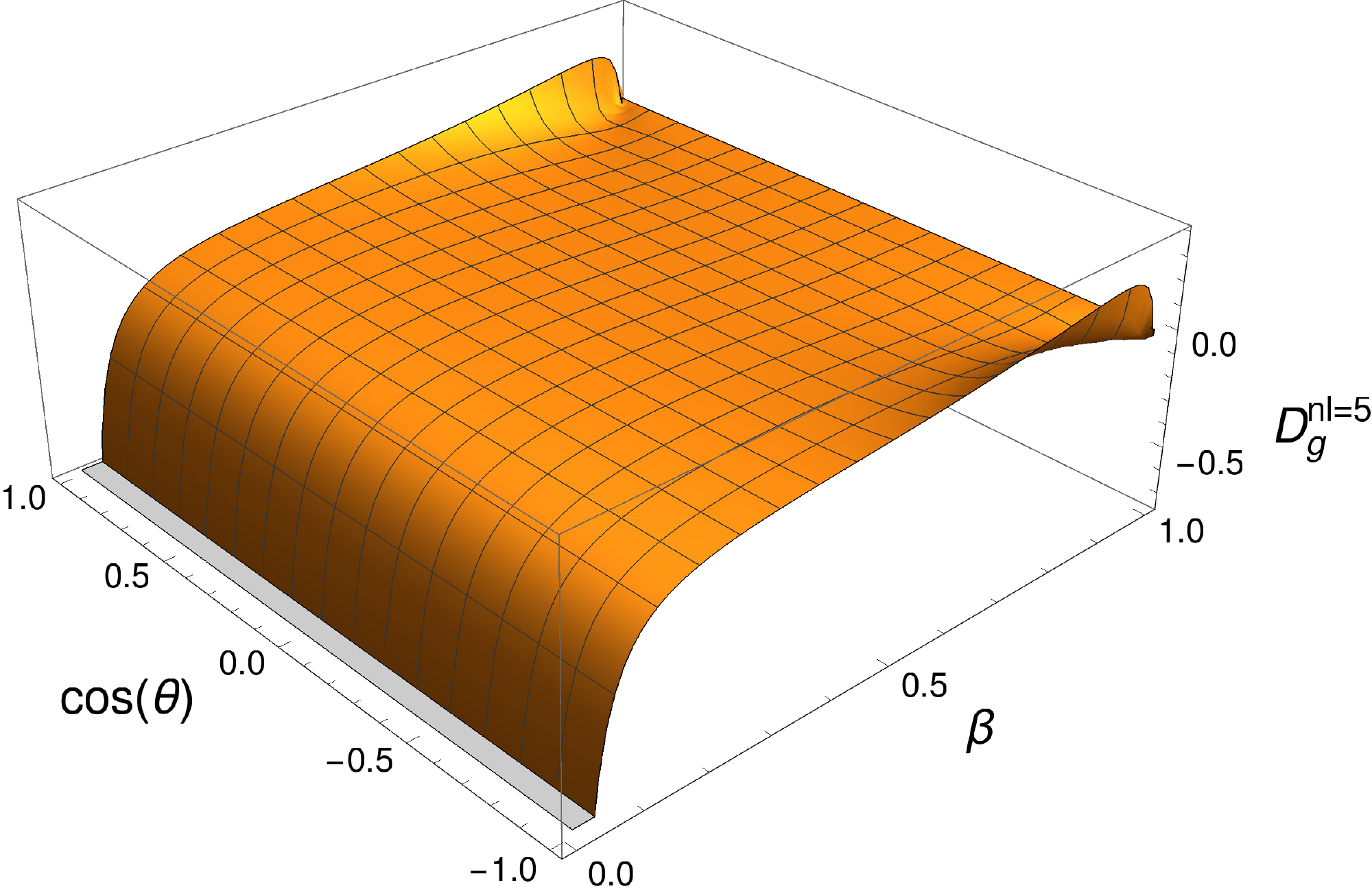}
              }%

\subfigure[]{
               \includegraphics[width=7.5cm]{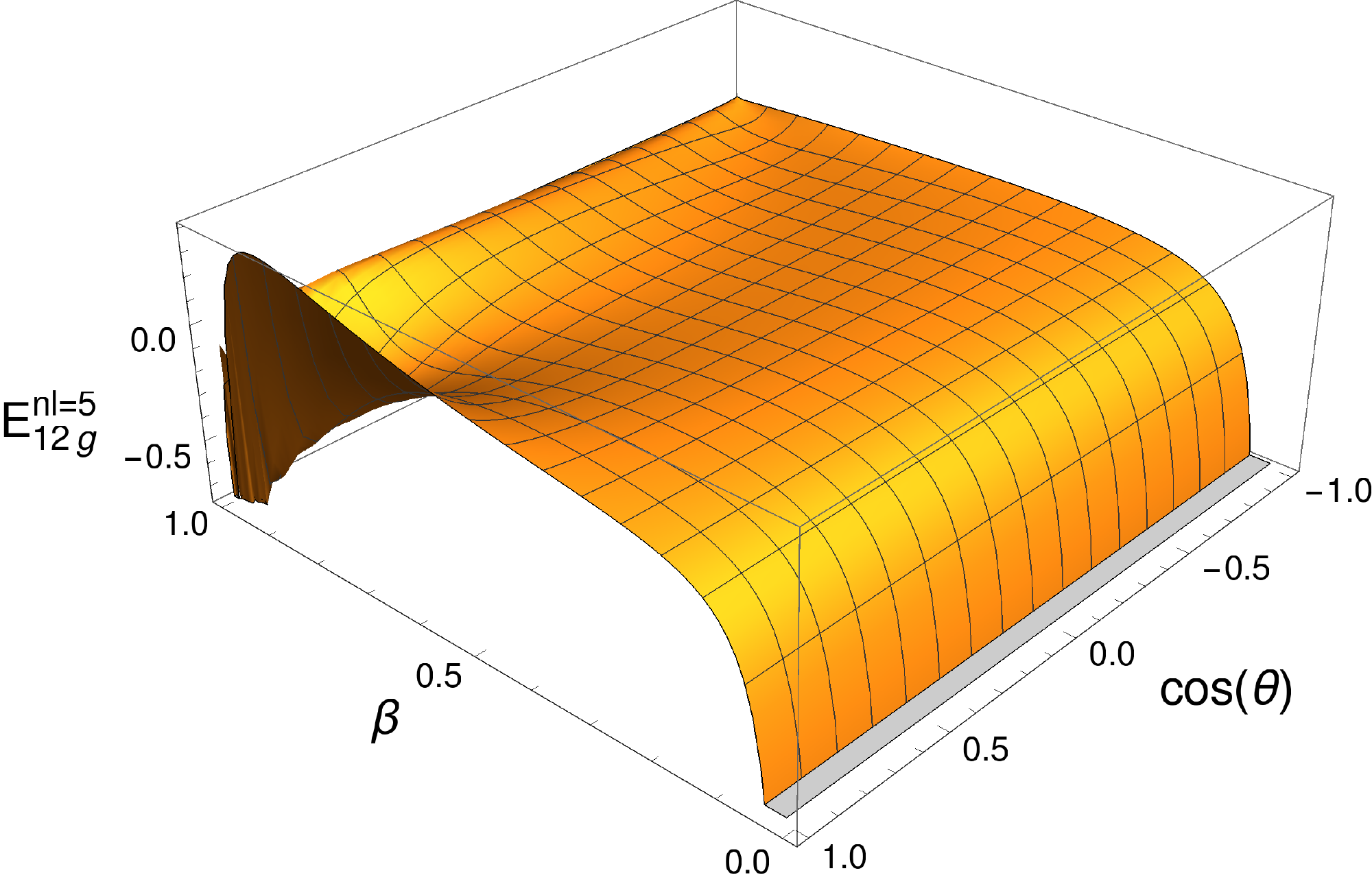}
              }%
\subfigure[]{
               \includegraphics[width=7.5cm]{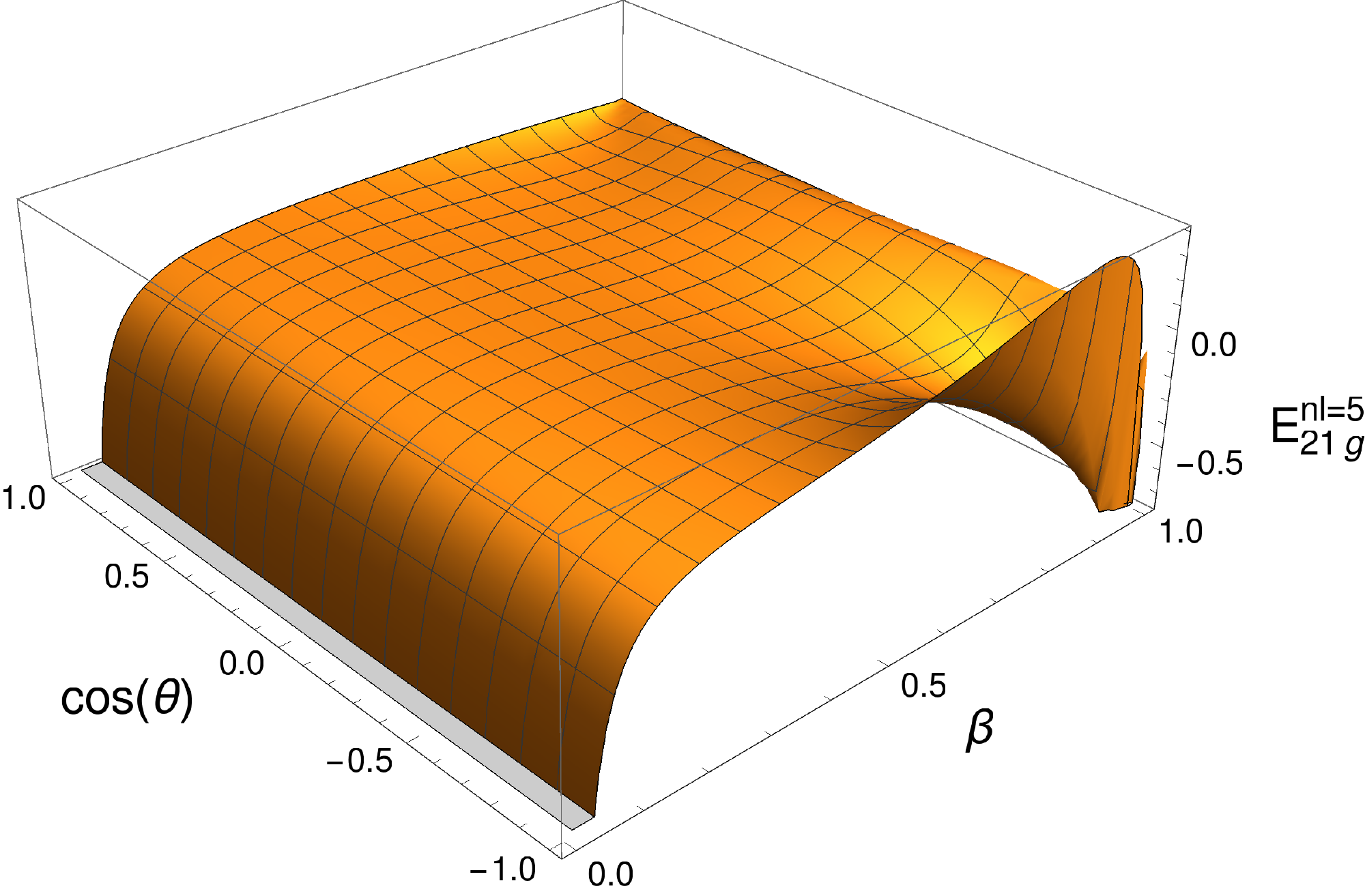}
              }%
\caption{Finite remainder coefficient functions of the spin-density
matrix in case of initial state gluons for $\nl=5$.}
\label{fig:sdmPlotgg}
\end{figure}

\begin{figure}[th!]
\subfigure[]{
               \includegraphics[width=7.5cm]{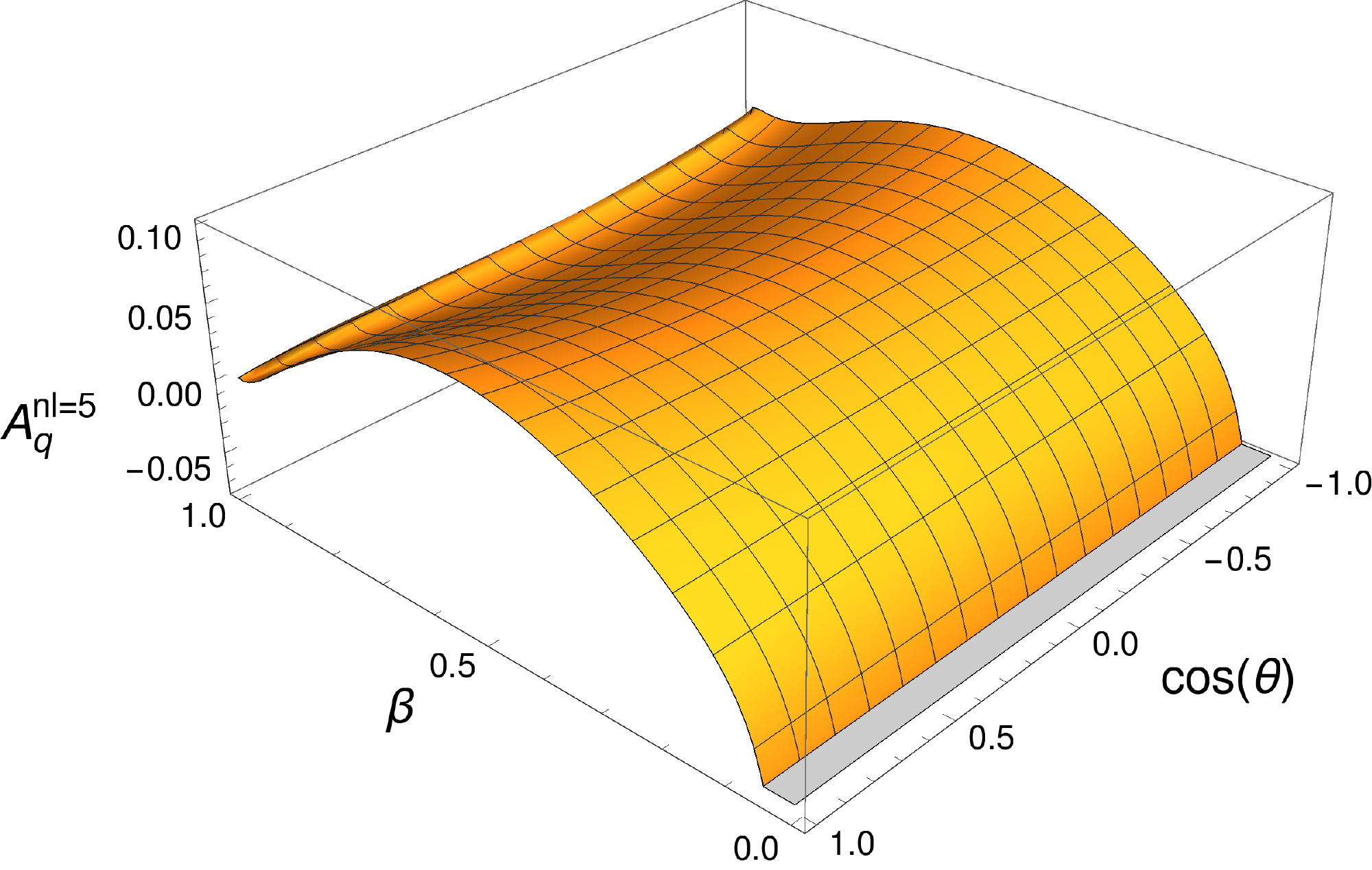}
              }%
\subfigure[]{
               \includegraphics[width=7.5cm]{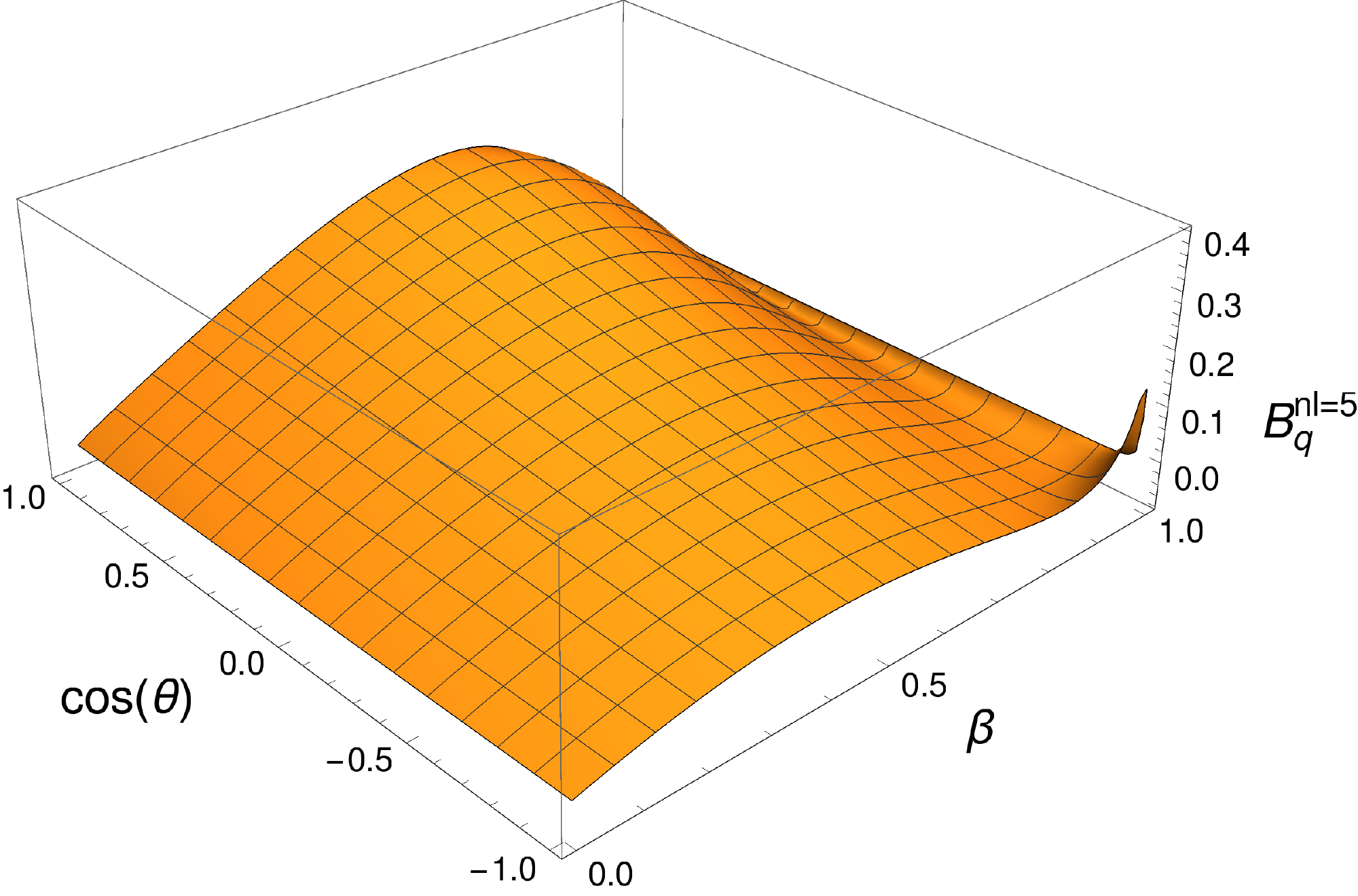}
              }%

\subfigure[]{
               \includegraphics[width=7.5cm]{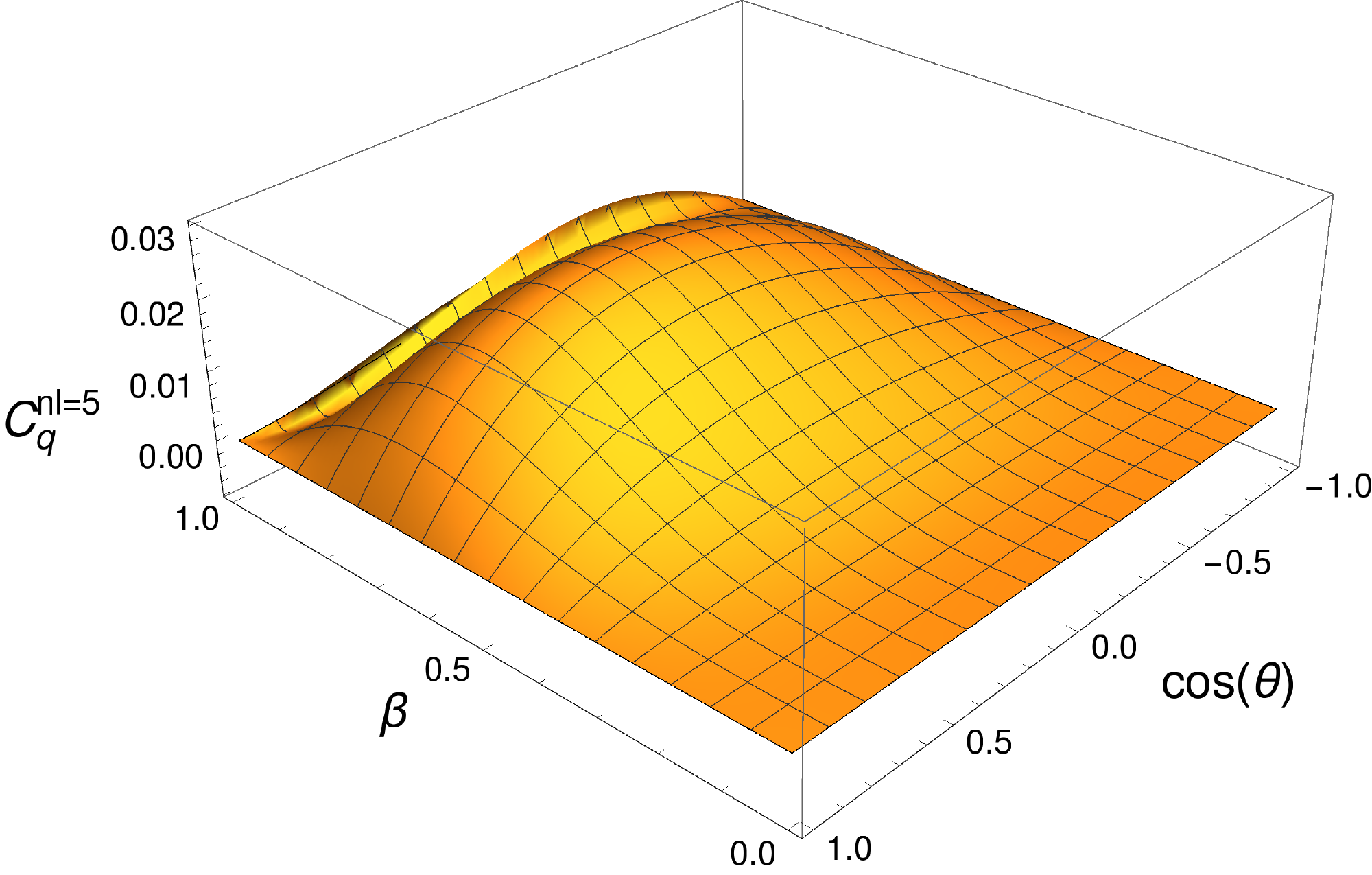}
              }%
\subfigure[]{
               \includegraphics[width=7.5cm]{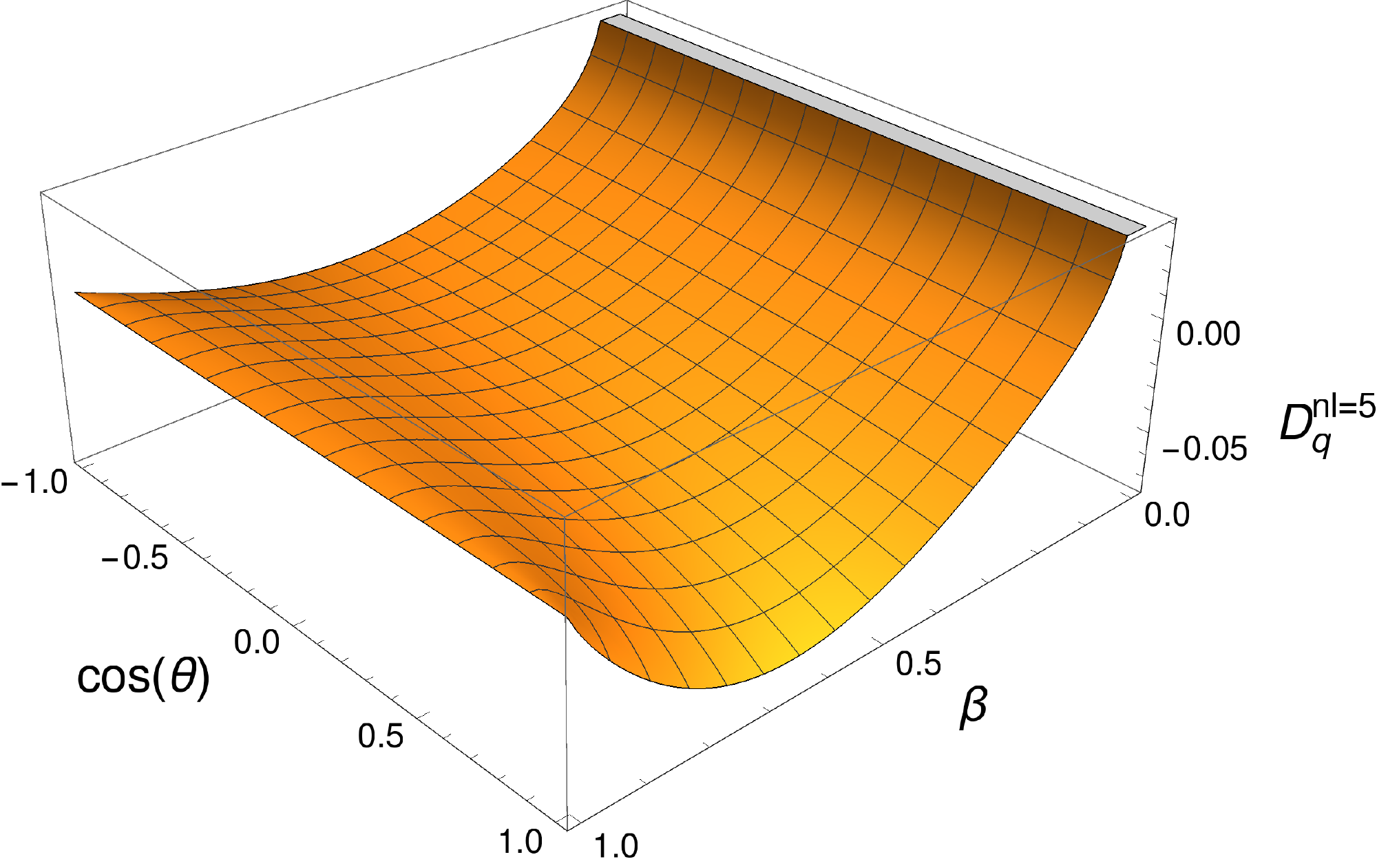}
              }%

\subfigure[]{
               \includegraphics[width=7.5cm]{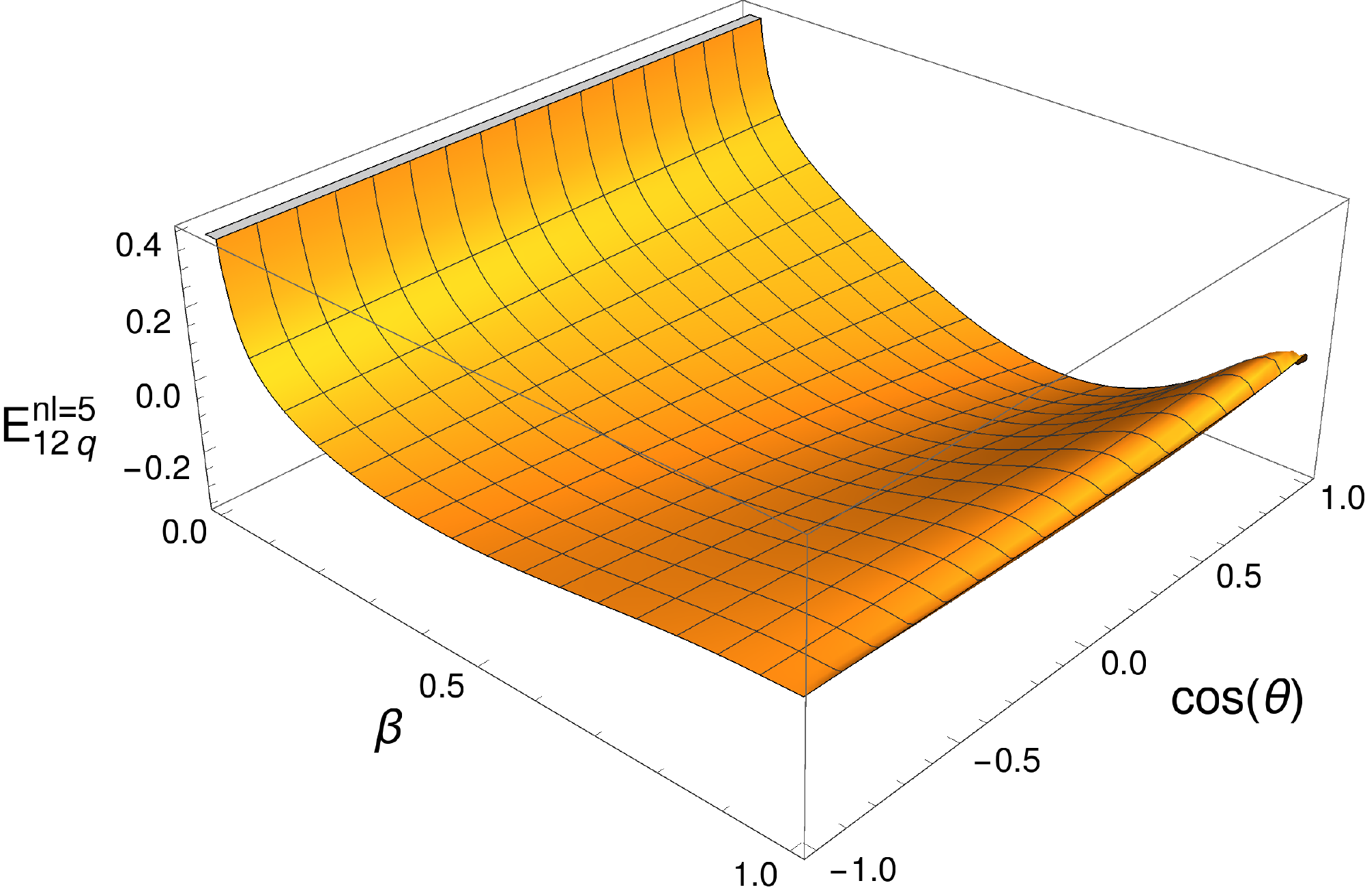}
              }%
\subfigure[]{
               \includegraphics[width=7.5cm]{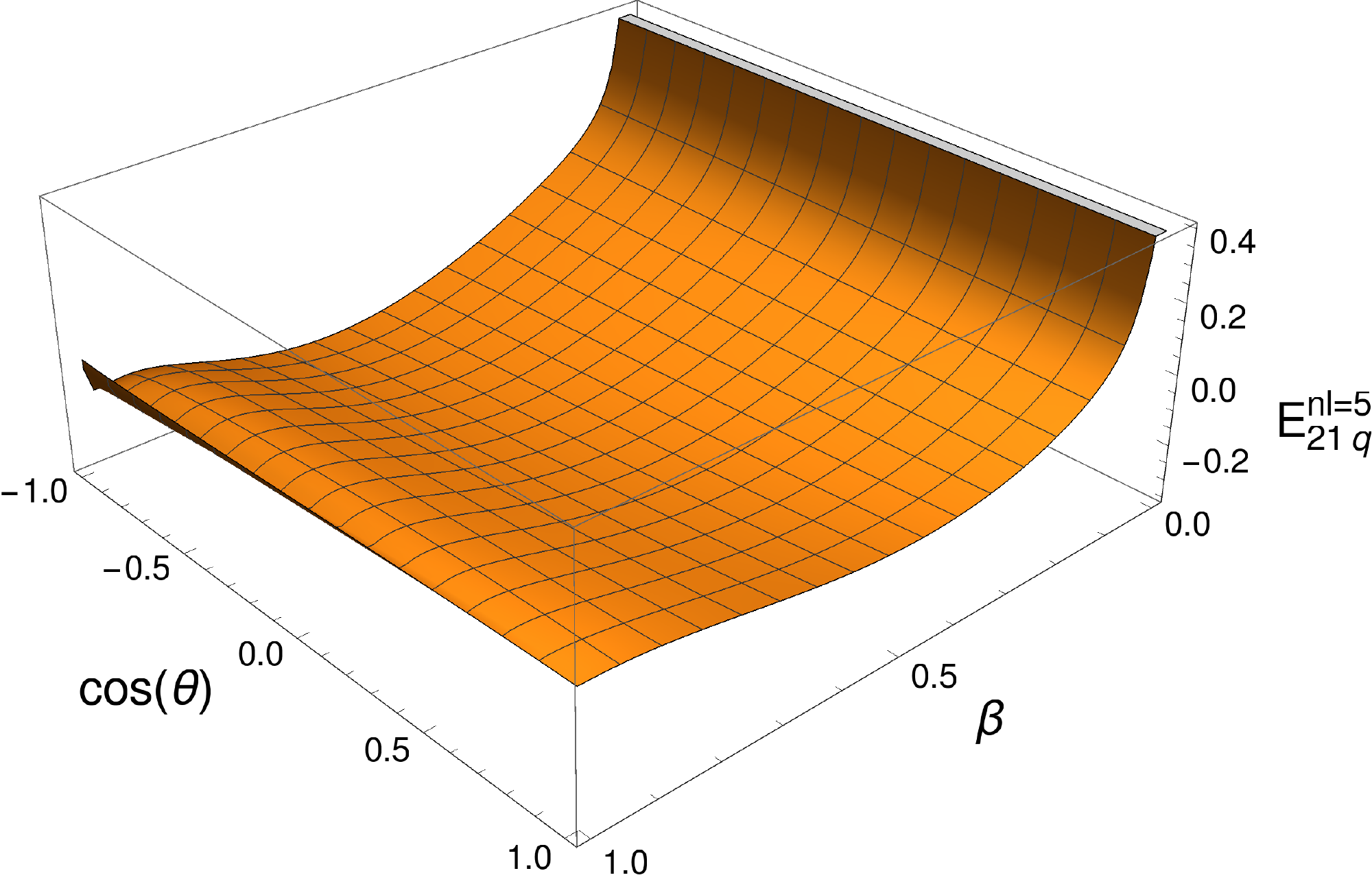}
              }%
\caption{Finite remainder coefficient functions of the spin-density
matrix in case of initial state quarks for $\nl=5$.}
\label{fig:sdmPlotqq}
\end{figure}

In this publication, we provide results for the finite remainders of all
coefficient functions. They are given in the form of an interpolation grid as
well as kinematic expansions in the high energy limit and near the production
threshold. The decomposition into the structure coefficients
yields maximal flexibility since they are
independent of the frame, definition of heavy quark spin vectors and other
conventions.

At tree-level we find the following non-vanishing coefficients in case of
gluons
\begin{align}
c^{(0)}_{11}&=\frac{-1}{x} \; ,  &c^{(0)}_{21}=\frac{-1}{1-x} \; , \\[0.2cm]
c^{(0)}_{15}&=\frac{2x-1}{x} \; , &c^{(0)}_{25}=\frac{2x-1}{1-x} \; , \\[0.2cm]
c^{(0)}_{17}&=c^{(0)}_{18}=-c^{(0)}_{11} \; ,
&c^{(0)}_{27}=c^{(0)}_{28}=-c^{(0)}_{21} \; .
\end{align}
These have the expected symmetry properties under the replacement $\cos
\theta\to-\cos\theta$ or $x\to1-x$. In case of quark-anti-quark
annihilation we find only
\begin{equation}
c^{(0)}_{13} =\frac{1}{2} \; , \qquad c^{(0)}_{23}=\frac{-1}{6} \; ,
\end{equation}
to be non-vanishing.
We find that in both cases, quark and gluon initial state, one spin structure 
has vanishing coefficients for every color structure at one and two-loops.
Indeed, in the case of gluons the coefficients $c_{i6}$ vanish, as do
the coefficients $c_{i4}$ in the case of quarks. All other spin
structures have non-vanishing coefficients.

\paragraph{The high energy limit} of the coefficients was calculated as an
analytic power-logarithmic
expansion in $m_s = \frac{m_t}{s}$ up to
$\order{m_s^4}$, using the boundary expressions for the master integrals.
This expansion assumes that $t,u \gg m_t^2$ and is therefore not valid
in the region of high-energy forward/backward scattering.
The results were cross-checked against the spin summed amplitude.

We want to mention that the depth of the expansion does not translate easily to
the expansion depth of the square summed or spin correlated matrix
element since there is a non trivial dependence on $m_s$
(or $\beta$ in case of the threshold expansion) hidden in
the spin structures themselves. 

\paragraph{The ``bulk'' region} is parameterized on a grid which is specified by equally spaced
points in $\beta$
\begin{align}
  \beta_i = i/80 \; , \qquad i \in [1,79] \; ,
\end{align}
and two additional points close to the high energy boundary.
The point $\beta_{80} = 0.999$ is sufficient for LHC with 8 TeV
center-of-mass energy, which was the extent of the interpolation grid
of the spin summed calculation. Here, we extend the grid to $\beta_{81} =
0.9997$ which corresponds to a center of mass energy of 14 TeV, for
contemporary applications. For $\cos\theta$ we choose 42 points
obtained from
\begin{align}
 \cos \theta = \pm x_i \; , \qquad i \in [1,21] \; ,
\end{align}
where we chose the $x_i$ as the 21 points obtained from the Gauss-Kronrod
integration rule of degree 10. Values for $\beta < 0.1$ were obtained from the
threshold expansion of the master integrals. The dependence on the number of
light fermions is kept.

In order to illustrate our results we plot the coefficients of the
spin density matrix. We introduce the following normalization factors,
which were also used for the presentation of the results in
\cite{Baernreuther:2013caa}
\begin{align}
  N_g = \frac{\beta(1-\beta^2)}{4096\pi}
  \;  \qquad \text{and}\; \qquad
  N_q = \frac{\beta(1-\beta^2)}{576\pi} \; ,
\end{align}
and define
\begin{align}
  \mathcal{R}_g^F = N_g 2 \Re
\braket{\mathcal{M}_g^0}{\mathcal{F}_g^2}(s_t,s_{\bar{t}}) \; , \\
  \mathcal{R}_q^F = N_q 2 \Re
\braket{\mathcal{M}_q^0}{\mathcal{F}_q^2}(s_t,s_{\bar{t}}) \; ,, 
\end{align}
which have the same decomposition as in Eq.~\eqref{eq:sdm-final}.
The coefficient functions of $\mathcal{R}_g^F$ and $\mathcal{R}_q^F$ 
are visualised in Figs.~\ref{fig:sdmPlotgg} and 
\ref{fig:sdmPlotqq} for $\nl=5$. 
 The function $A^{\nl=5}_{q,g}$ is 
the spin-summed and averaged two-loop finite remainder and was checked
against the result from \cite{Baernreuther:2013caa}. The
function $B^{\nl=5}_{q,g}$ describes the transverse polarization of the top quarks
resulting from absorptive parts of the amplitude. At tree-level, this
coefficient vanishes due to the absence of complex couplings in
QCD. At higher orders, the non-vanishing imaginary part of the virtual
amplitudes yields non-zero coefficients. The remaining functions encode the spin correlations between the top and anti-top
quark since their structures contain both spin vectors. In the gluon
channel, all expected symmetry properties under
$\cos\theta\to-\cos\theta$ of the coefficient functions are clearly fulfilled.

\begin{figure}
  \centering
  \includegraphics[width=12cm]{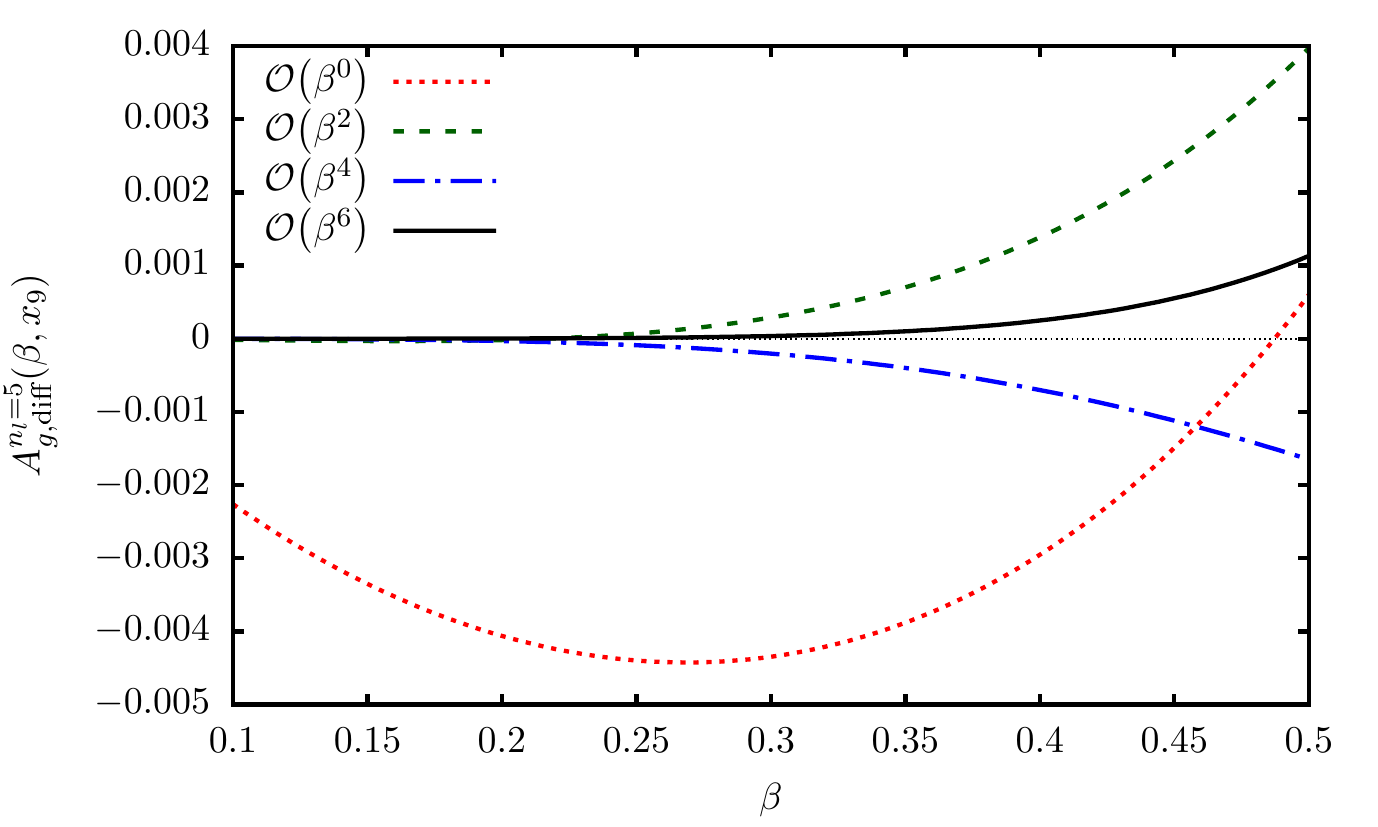}
  \caption{The difference between the threshold expansion for
           coefficient $A_g$ up to $\beta^n$ with $n=0,2,4,6$ and results
           from numerical integration for a fixed angle $\theta$.}
  \label{fig:thresA}
\end{figure}
 
\begin{figure}
\subfigure[]{ 
\includegraphics[width=7.5cm]{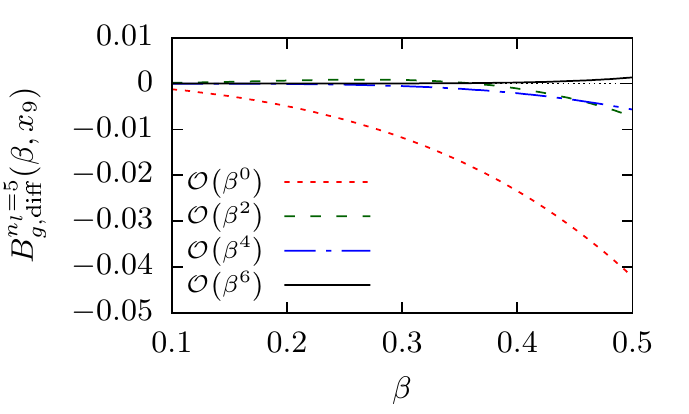}
}%
\subfigure[]{ 
\includegraphics[width=7.5cm]{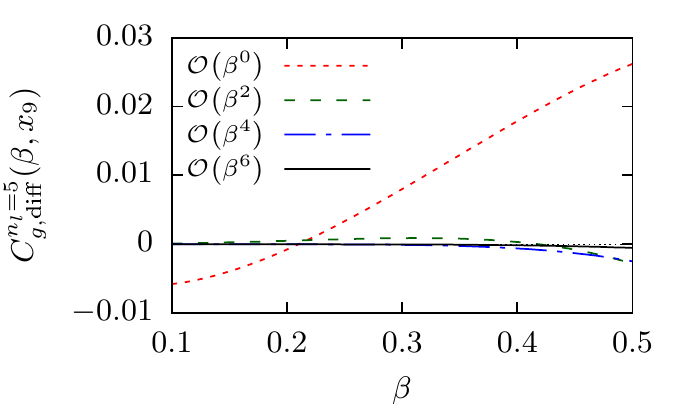}
}
\subfigure[]{ 
\includegraphics[width=7.5cm]{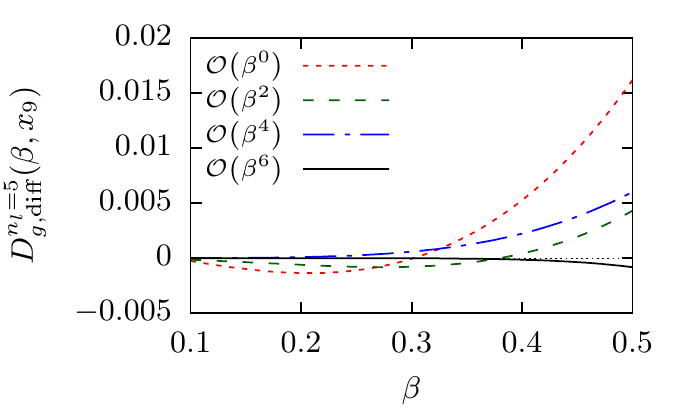}
}%
\subfigure[]{ 
\includegraphics[width=7.5cm]{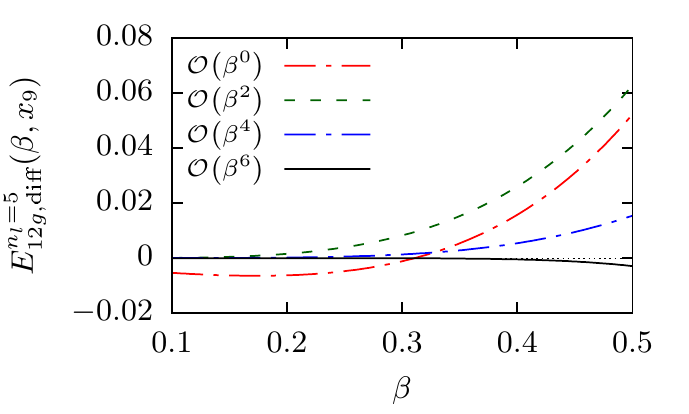}
}
  \caption{The difference between the threshold expansion for
           coefficient $B_g,C_g,D_g,E_{12g}$ up to $\beta^n$ with $n=0,2,4,6$ and results
           from numerical integration for a fixed angle $\theta$.}
 \label{fig:thresSpin}
\end{figure}

\paragraph{The threshold region} is covered by points obtained from the deep
power-logarithmic expansions of the master integrals. In addition we perform
a power-log expansion in $\beta$ for all coefficients up to $\beta^2$.
This is done for different but fixed scattering angles $\cos\theta$
\begin{align*}
  c_{ij}(\beta,\cos \theta_n) = \sum_{k=-2}^{2} \sum_{l=0}^2
\tilde{c}_{ij,kl,n}\beta^k\ln^l\beta
\end{align*}
The dependence on $\theta$ was recovered by performing a fit for each set
$\{\tilde{c}_{ij,kl,n}\}_n$ to a polynomial
$\tilde{c}_{ij,kl} = \sum_{n=0}^{2+k} a_n \cos^n\theta$ separately for the real and imaginary
part. The results are also available in electronic format together
with this paper. This expansion was used to determine the
corresponding coefficients of the spin density matrix as well.
Up to $\order{\beta^0}$ we reproduce the analytic result obtained for the
spin-summed case in \cite{Baernreuther:2013caa}.

To study the quality and convergence of the expansion, we also
calculated the density matrix for a fixed angle (chosen to be the
point $x_9$) up to order $\order{\beta^6}$. In Figs.~\ref{fig:thresA} and
\ref{fig:thresSpin} we compare this expansion against the results
obtained from the interpolation grid. We show the
difference\footnote{We do not plot the relative difference, because
  the coefficient functions have a zero in the plotted region.}
\begin{align}
  \left(X^{\nl=5}_{\rm diff}\right) (\beta,x_9) = 
     \left(X^{\nl=5}_{\rm thres}\right) (\beta,x_9)-
     \left(X^{\nl=5}_{\rm grid}\right) (\beta,x_9) \; ,
\end{align}
with $X \in \{A_g,B_g,C_g,D_g,E_{12g}\}$ for different expansion depths of
$\left(X^{\nl=5}_{\rm thres}\right) (\beta,x_9)$.
The series seems to converge nicely and, if expanded up to $\order{\beta^6}$,
provides a reasonable description of the amplitude in the region $\beta<0.3$.

\section{Conclusions and Outlook}
We presented the decomposition of the heavy-quark pair production amplitude in
terms of spin and color structures at two-loop level. We provide results in
terms of interpolation grids and kinematic expansions for these coefficients.
As a first application we calculated the spin-density matrix for top-quark
pairs.
With this work we provide the missing piece needed for the calculation of
on-shell top-quark pair production and decay at NNLO in QCD including
spin-correlation effects in the narrow width approximation. We
improved the numerical results obtained for the involved master
integrals by changing to a partly canonical basis. The 
incorporation of these amplitudes in a full-fledged calculation of top-quark
pair production and decay is work in progress.

The full set of results of this paper is available at: \\
\href{https://git.rwth-aachen.de/mczakon/PolarizedTTNNLO}{https://git.rwth-aachen.de/mczakon/PolarizedTTNNLO}.

\begin{acknowledgments}
L. Chen acknowledges support by a scholarship from the China
Scholarship Council (CSC). R. Poncelet was supported by the Deutsche
Forschungsgemeinschaft through Graduiertenkolleg GRK 1675, and in part
by the German Research Foundation (DFG) under Grant No. WO 1900/2.
\end{acknowledgments}

\appendix

\section{Renormalization constants and anomalous dimensions}
\label{sec:RConst}

We list all necessary renormalization constants up to the needed power in $\ep$.
The on-shell renormalization constants are
\begin{eqnarray}
Z_g &=& 1
+ \left(\frac{\asnf}{2\pi}\right) \tf \nh
\biggl\{
  - {2 \over 3 \ep}
  - {2 \over 3} \lmu{}
  - {1 \over 3} \ep \lmu2
  - {\pi^2 \over 18} \ep
  - {1 \over 9} \ep^2 \lmu3
  - {\pi^2 \over 18} \ep^2 \lmu{}
  + {2 \over 9} \ep^2 \z3
\biggr\}
\nonumber\\
& &
+ \left(\frac{\asnf}{2\pi}\right)^2 \tf \nh
\biggl\{
  \tf \nh \biggl[
      {4 \over 9 \ep} \lmu{}
    + {2 \over 3} \lmu2
    + {\pi^2 \over 27}
  \biggr]
+ \tf \nl \biggl[
    - {4 \over 9 \ep^2}
    - {4 \over 9 \ep} \lmu{}
    - {2 \over 9} \lmu2
    - {\pi^2 \over 27}
  \biggr]
\nonumber\\
& &
+ \cf \biggl[
    - {1 \over 2 \ep}
    - \lmu{}
    - {15 \over 4}
  \biggr]
+ \ca \biggl[
      {35 \over 36 \ep^2}
    + {13 \over 18 \ep} \lmu{}
    - {5 \over 8 \ep}
    - {5 \over 4} \lmu{}
    + {1 \over 9} \lmu2
    + {13 \over 48}
    + {13 \pi^2 \over 216}
  \biggr]
\biggr\}\, ,
\nonumber \\ \nonumber \\
Z_q &=& 1 + \left(\frac{\asnf}{2\pi}\right)^2 \cf \tf \nh
\biggl[
  \frac{1}{4\ep}
  + \frac{1}{2} \lmu{}
  - \frac{5}{24}
\biggr]\, ,
\nonumber \\ \nonumber \\
Z_Q &=& 1
+ \left(\frac{\asnf}{2\pi}\right) \cf
\biggl\{
- \frac{3}{2\ep}
- 2
- \frac{3}{2}\lmu{}
- 4 \ep
- 2 \ep \lmu{}
- \frac{3}{4} \ep \lmu2
- \frac{\pi^2}{8} \ep
- 8 \ep^2
- 4 \ep^2 \lmu{}
- \ep^2 \lmu2
\nonumber\\
& &
- \frac{1}{4} \ep^2 \lmu3
- \frac{\pi^2}{6} \ep^2
- \frac{\pi^2}{8} \ep^2 \lmu{}
+ \frac{1}{2} \ep^2 \z3
\biggr\}
+ \left(\frac{\asnf}{2\pi}\right)^2 \cf
\biggl\{
\tf \nh \biggl[
\frac{1}{4\ep}
+ \frac{1}{\ep} \lmu{}
+ \frac{947}{72}
+ \frac{11}{6} \lmu{}
\nonumber\\
& &
+ \frac{3}{2} \lmu2
- \frac{5\pi^2}{4}
\biggr]
+ \tf \nl \biggl[
- \frac{1}{2\ep^2}
+ \frac{11}{12\ep}
+ \frac{113}{24}
+ \frac{19}{6} \lmu{}
+ \frac{1}{2} \lmu2
+ \frac{\pi^2}{3}
\biggr]
+ \cf \biggl[
\frac{9}{8\ep^2}
+ \frac{51}{16\ep}
\nonumber\\
& &
+ \frac{9}{4\ep} \lmu{}
+ \frac{433}{32}
+ \frac{51}{8} \lmu{}
+ \frac{9}{4} \lmu2
- \frac{49\pi^2}{16}
+ 4 \lt1 \pi^2
- 6 \z3
\biggr]
+ \ca \biggl[
\frac{11}{8\ep^2}
- \frac{127}{48\ep}
- \frac{1705}{96}
\nonumber\\
& &
- \frac{215}{24} \lmu{}
- \frac{11}{8} \lmu2
+ \frac{5\pi^2}{4}
- 2 \lt1 \pi^2
+3 \z3
\biggr] 
\biggr\}
\, ,
\nonumber \\ \nonumber \\
Z_m &=& 1
+ \left(\frac{\asnf}{2\pi}\right) \cf
\biggl\{
- \frac{3}{2\ep}
- 2
- \frac{3}{2}\lmu{}
- 4 \ep
- 2 \ep \lmu{}
- \frac{3}{4} \ep \lmu2
- \frac{\pi^2}{8} \ep
- 8 \ep^2
- 4 \ep^2 \lmu{}
- \ep^2 \lmu2
\nonumber\\
& &
- \frac{1}{4} \ep^2 \lmu3
- \frac{\pi^2}{6} \ep^2
- \frac{\pi^2}{8} \ep^2 \lmu{}
+ \frac{1}{2} \ep^2 \z3
\biggr\}
+ \left(\frac{\asnf}{2\pi}\right)^2 \cf
\biggl\{
\tf \nh \biggl[
- \frac{1}{2\ep^2}
+ \frac{5}{12\ep}
+ \frac{143}{24}
\nonumber\\
& &
+ \frac{13}{6} \lmu{}
+ \frac{1}{2} \lmu2
- \frac{2\pi^2}{3}
\biggr]
+ \tf \nl \biggl[
- \frac{1}{2\ep^2}
+ \frac{5}{12\ep}
+ \frac{71}{24}
+ \frac{13}{6} \lmu{}
+ \frac{1}{2} \lmu2
+ \frac{\pi^2}{3}
\biggr]
\nonumber\\
& &
+ \cf \biggl[
\frac{9}{8\ep^2}
+ \frac{45}{16\ep}
+ \frac{9}{4\ep} \lmu{}
+ \frac{199}{32}
+ \frac{45}{8} \lmu{}
+ \frac{9}{4} \lmu2
- \frac{17\pi^2}{16}
+ 2 \lt1 \pi^2
- 3 \z3
\biggr]
\nonumber\\
& &
+ \ca \biggl[
\frac{11}{8\ep^2}
- \frac{97}{48\ep}
- \frac{1111}{96}
- \frac{185}{24} \lmu{}
- \frac{11}{8} \lmu2
+ \frac{\pi^2}{3}
- \lt1 \pi^2
+ \frac{3}{2} \z3
\biggr] 
\biggr\}
\, ,
\end{eqnarray}
where $\lmu{} = \ln \mu^2/m^2$. The on-shell
wave-function renormalization constants for the gluon and light quark
fields have been taken from \cite{Czakon:2007ej, Czakon:2007wk}.

For the heavy-quark wave-function and mass
renormalization constants we used expressions from \cite{Broadhurst:1991fy}.
The $\MSbar$ renormalization constant for the strong coupling up
to $\order{\asnf^2}$ is given in terms of beta-function coefficients
\begin{eqnarray}
Z_\as =
   1
   - \left( {\asnf \over 2 \pi} \right) \frac{b_0}{2 \epsilon} 
   +  \left( {\asnf \over 2 \pi} \right)^2 \left(
     \frac{b_0^2}{4\epsilon^2}
     - \frac{b_1}{8 \epsilon}
     \right)
\, ,
\end{eqnarray}
where
\begin{eqnarray}
b_0 = {11 \over 3} \ca - {4 \over 3} \tf \nf \, ,
\qquad
b_1 = {34 \over 3} \ca^2 - {20 \over 3} \ca \tf \nf - 4 \cf \tf \nf \, .
\end{eqnarray}
The two-loop decoupling constant for the strong coupling is given by
\cite{Bernreuther:1981sg}
\begin{eqnarray}
\zeta_\as &=& 1
+ \left(\frac{\asnl}{2\pi}\right) \tf \nh
\biggl\{
\frac{2}{3} \lmu{}
+ \frac{1}{3} \ep \lmu2
+ \frac{\pi^2}{18} \ep
+ \frac{1}{9} \ep^2 \lmu3
+ \frac{\pi^2}{18} \ep^2 \lmu{}
- \frac{2}{9} \ep^2 \z3
\biggr\}
\nonumber\\
& &
+ \left(\frac{\asnl}{2\pi}\right)^2 \tf \nh
\biggl\{
\frac{4}{9} \tf \nh
\lmu2  
+ \cf \biggl[
\frac{15}{4}
+ \lmu{}
\biggr]
+ \ca \biggl[
- \frac{8}{9}
+ \frac{5}{3} \lmu{}
\biggr]
\biggr\}\, .
\end{eqnarray}
The anomalous dimension of the $\mathbf{Z}$ operator used to define the finite
remainder function is given by \cite{Ferroglia:2009ii}
\begin{equation}\label{eq:IRGamma}
\begin{split}
{\bf \Gamma}_{\cal M}(\{\underline{p}\},\{\underline{m}\},\mu)\,
& = \,\sum\limits_{(i,j)}\frac{{\bf T}_i\cdot{\bf
    T}_j}{2}\,\gamma_{\text{cusp}}\left(\asnl\right)\,
\ln\frac{\mu^2}{-s_{ij}}\,+\,\sum\limits_i
\gamma^i\left(\asnl\right)\\ 
& \hspace{-2.5cm} -\,\sum\limits_{(I,J)}\frac{{\bf T}_I\cdot{\bf T}_J}{2}\,
\gamma_{\text{cusp}}\left(\beta_{IJ},\asnl\right)\,+\,\sum\limits_I
\gamma^I\left(\asnl\right)\,
+\,\sum\limits_{I,j}{\bf T}_I\cdot{\bf
  T}_j\,\gamma_{\text{cusp}}\left(\asnl\right)\,
\ln\frac{m_I\,\mu}{-s_{Ij}}\\
& \hspace{-2.5cm} +\,\sum\limits_{(I,J,K)}i\,f^{abc}\,{\bf T}_I^a\,{\bf T}_J^b\,
{\bf T}_K^c\,F_1(\beta_{IJ},\beta_{JK},\beta_{KI})\\ 
& \hspace{-2.5cm} +\,\sum\limits_{(I,J)}\sum\limits_k\,i\,f^{abc}\,
{\bf T}_I^a\,{\bf T}_J^b\,{\bf T}_k^c\,f_2\left(\beta_{IJ},
\ln\frac{-\sigma_{Jk}\,v_J\cdot p_k}{-\sigma_{Ik}\,
v_I\cdot p_k}\right)
\; .
\end{split}
\end{equation} 
The lower case indices denote sums over massless particles while capital
letters denote sums over massive particles. The brackets $(i,j,...)$ indicate
that the sums go over different indicies. The action of the color operator ${\bf
T}^a_i$ dependence on the type of the parton with index $c$ it acts on.
After projecting the result onto the index $b$ we have that
in case of a gluon  $({\bf T}^a)_{bc} = -if^{abc}$ .
In case of an outgoing quark (incoming
anti-quark) $({\bf T}^a)_{bc} =T^a_{bc} $ and
$({\bf T}^a)_{bc} = - T^a_{bc} $ for a incoming quarks (or outgoing
anti-quark). For the kinematic dependence we have the definitions
\begin{align*}
s_{ij} = 2\sigma_{ij}p_ip_j+i0^+ \;\;\;&\text{with}\;\;\; \sigma_{ij}=+1
\;\;\text{if } p_i \text{ and } p_j\text{ in/out going and } \sigma_{ij}=-1
\;\text{otherwise}\\
&p^2_I=m^2_I\;\;\;v_I=p_I/m_I\;\;\cosh \beta_{IJ} = -s_{IJ}/2m_Im_J\;. 
\end{align*} 
In contrast to the spin and color summed case, where all triple color
correlators vanish \cite{Czakon:2013hxa}, they are essential for the infrared finiteness of the
structure coefficients.
We also list the anomalous dimensions occurring in
Eq.~\eqref{eq:IRGamma} necessary to obtain the finite remainders of
the two-loop amplitudes. The anomalous dimensions related to a single
parton (collinear in origin for massless partons and soft in origin
for massive partons) are \cite{Becher:2009qa, Becher:2009kw}
\begin{eqnarray}
\gamma^g\left( \asnl \right) &=& \left( \frac{\asnl}{2\pi} \right) 
\biggl\{
- \frac{11}{6} \ca
+ \frac{2}{3} \tf \nl
\biggr\}
+ \left( \frac{\asnl}{2\pi} \right)^2
\biggl\{
\ca^2 \biggr[
- \frac{173}{27}
+ \frac{11\pi^2}{72}
+ \frac{1}{2} \z3
\biggr]
\nonumber \\
& & 
+ \ca \tf \nl \biggl[
\frac{64}{27}
- \frac{\pi^2}{18}
\biggr]
+ \cf \tf \nl
\biggr\} \, ,
\\ \nonumber \\
\gamma^q\left( \asnl \right) &=& - \left( \frac{\asnl}{2\pi} \right)
\frac{3}{2} \cf
+ \left( \frac{\asnl}{2\pi} \right)^2 \cf
\biggl\{
\ca \biggr[
- \frac{961}{216}
- \frac{11\pi^2}{24}
+ \frac{13}{2} \z3
\biggr]
\nonumber \\
& & 
+ \cf \biggl[
- \frac{3}{8}
+ \frac{\pi^2}{2}
- 6 \z3
\biggr]
+ \tf \nl \biggl[
\frac{65}{54}
+ \frac{\pi^2}{6}
\biggr]
\biggr\} \, ,
\\ \nonumber \\
\gamma^Q\left( \asnl \right) &=& - \left( \frac{\asnl}{2\pi} \right) \cf
+ \left( \frac{\asnl}{2\pi} \right)^2 \cf
\biggl\{
\ca \biggr[
- \frac{49}{18}
+ \frac{\pi^2}{6}
- \z3
\biggr]
+ \frac{10}{9} \tf \nl
\biggr\} \, .
\end{eqnarray}
The cusp anomalous dimensions are given by \cite{Korchemsky:1987wg,
  Kidonakis:2009ev}
\begin{eqnarray}
\gamma_{\rm cusp}\left( \asnl \right) &=& \frac{\asnl}{\pi}
+\left( \frac{\asnl}{2\pi} \right)^2
\biggl\{
\ca \biggl[
\frac{67}{9}
- \frac{\pi^2}{3}
\biggr]
- \frac{20}{9} \tf \nl
\biggr\} \, ,
\\ \nonumber \\
\gamma_{\rm cusp}\left( \beta, \asnl \right) &=&
\gamma_{\rm cusp}\left( \asnl \right) \beta \coth \beta
\nonumber \\
& + &
\left( \frac{\asnl}{2\pi} \right)^2 2 \ca
\biggl\{
\coth^2\beta
\biggl[
{\rm Li}_3(e^{-2\beta})
+ \beta \, {\rm Li}_2(e^{-2\beta})
- \z3
+ \frac{\pi^2}{6} \beta
+ \frac{1}{3} \beta^3
\biggr]
\nonumber \\
& &
+ \coth\beta
\biggl[
{\rm Li}_2(e^{-2\beta})
-2\beta \ln (1 - e^{-2\beta})
- \frac{\pi^2}{6}(1 + \beta)
- \beta^2
- \frac{1}{3} \beta^3
\biggr]
\nonumber \\
& &
+ \frac{\pi^2}{6}
+ \z3
+ \beta^2
\biggr\} \, .
\end{eqnarray}
The two functions $F_1$ and $f_2$ are given by
\begin{align}
  F_1(\beta_{12},\beta_{23},\beta_{31})  = \frac{1}{3}\sum_{I,J,K}^3 \ep_{I,J,K}
     \frac{\as}{4\pi}g(\beta_{IJ}) \gamma_{\rm cusp}(\beta_{KI},\as)
  \; , \\
  f_2\left(\beta_{12},\ln \frac{-\sigma_{23}v_2p_3}{-\sigma_{13} v_1 p_3}\right)
    = -\frac{\as}{3\pi} g(\beta_{12})\gamma_{\rm cusp}(\as)
  \ln\left(\frac{-\sigma_{23}v_2p_3}{-\sigma_{13} v_1 p_3}\right) \; ,
\end{align}
with the function
\begin{align}
g(\beta) = \coth \beta
\left[\beta^2+2\beta\ln(1-e^{-2\beta})-\text{Li}_2(e^{-2\beta})+\frac{\pi^2}{6} \right]
 -\beta^2-\frac{\pi^2}{6} \; .
\end{align}


\end{document}